# Phycobilisome core architecture influences photoprotective quenching by the Orange Carotenoid Protein

Ayesha Ejaz[1], Markus Sutter[2,3,4], Sigal Lechno-Yossef[2], Cheryl A. Kerfeld[2,3,4,5], Allison Squires[6,7,8]*

[1] Department of Chemistry, University of Chicago, Chicago, IL

[2] MSU-DOE Plant Research Laboratory, Michigan State University, East Lansing, MI, USA

[3] Environmental Genomics and Systems Biology Division, Lawrence Berkeley National Laboratory, Berkeley, CA, USA

[4] Molecular Biophysics and Integrated Bioimaging Division, Lawrence Berkeley National Laboratory, Berkeley, CA, USA

[5] Department of Biochemistry and Molecular Biology, Michigan State University, East Lansing, MI, USA

[6] Pritzker School of Molecular Engineering, University of Chicago, Chicago, IL, USA

[7] Institute for Biophysical Dynamics, University of Chicago, Chicago, IL, USA

[8] Chan-Zuckerberg Biohub Chicago, Chicago, IL, USA

* e-mail correspondence: asquires@uchicago.edu

# Abstract

Photosynthetic organisms rely on sophisticated photoprotective mechanisms to prevent oxidative damage under high or fluctuating solar illumination. Cyanobacteria, which have evolved a modular, water-soluble light harvesting complex—the phycobilisome—achieve photoprotection through a unique, photoactivatable quencher called the Orange Carotenoid Protein (OCP). Although phycobiliproteins are highly conserved, phycobilisomes take on different macromolecular architectures in different species of cyanobacteria, and it is not well understood whether or how these structures relate to changes in photoprotective function. For example, although two binding sites for dimers of OCP were recently discovered on the tricylindrical phycobilisome of *Synechocystis* sp. PCC 6803, these sites appear to be inaccessible on a pentacylindrical architecture (*Anabaena* sp. PCC 7120). Therefore, to learn whether OCP functions similarly across species with different core architectures, we experimentally compare the photophysical states accessible to prototypical tricylindrical and pentacylindrical phycobilisomes, with and without OCP, at the single-molecule level using an Anti-Brownian ELectrokinetic (ABEL) trap. We compare our data to Monte Carlo simulations of exciton transfer in compartmental models of phycobilisomes with OCP bound at different combinations of predicted docking sites. Our results suggest that while some aspects of OCP function are influenced by phycobilisome architecture, others are surprisingly well-conserved: OCP appears to bind at different locations in each architecture and cross-species OCP-phycobilisome compatibility is asymmetric, yet the quenching strength and dimeric binding of OCP appear to be similar for both phycobilisome architectures. Together, our findings provide new insights into how the uniquely modular architecture of phycobilisomes enables robust conservation as well as fine-tuning of the OCP quenching mechanism across species.

# Introduction

Sunlight drives photosynthesis and solar cells alike, but high light can create damage. Among light-harvesting organisms, cyanobacteria have evolved a unique light-harvesting antenna complex, the phycobilisome (1–4), and an accompanying capability for non-photochemical quenching by the photoswitchable Orange Carotenoid Protein (OCP) (5, 6), that together have enabled survival across a wide range of environments and solar illumination. Both phycobilisomes and OCP appear to be modular, with diverse architectures of phycobilisomes assembled from similar subunits in different species, and multiple paralogs of OCP present in many species. Although recently binding sites for OCP were discovered on the phycobilisome of one species, the underlying principles that govern quenching of these massive phycobilisomes (>6 MDa) by OCP (~34 kDa) across different species of cyanobacteria remain poorly understood.

A phycobilisome structure can be as simple as a cluster of rods or take on a more elaborate rod-core architecture with 2-5 core cylinders and 6 or more rods (7). The phycobilisome funnels energy from the more distal, blue-shifted pigments towards red-shifted pigments in the core, from which energy is transferred to a photosystem in the thylakoid membrane (8, 9). OCP is present in most phycobilisome-containing cyanobacteria (10, 11), including cyanobacteria with phycobilisome architectures that have tricylindrical or pentacylindrical cores (**Fig. 1a-b**). Phycobilisomes from *Synechocystis* sp. PCC 6803 (denoted throughout as tri-PBS) take on a prototypical tricylindrical core structure, where six rods composed of C-phycocyanin (Cpc) radiate outwards from an allophycocyanin (Apc) core containing three stacked cylinders. The rods comprise three Cpc hexamers each and the core cylinders each contain four Apc trimers, all linked together via colorless linker proteins. In contrast, a prototypical pentacylindrical phycobilisome from *Anabaena* sp. PCC 7120 (denoted throughout as penta-PBS) has two extra Apc half-cylinders in the core and eight Cpc rods (12, 13). The sequences of Apc for tri-PBS and penta-PBS are 83% identical (**SI Appendix Fig. S1**). In total, the established structure for tri-PBS contains 320 protein chains and 396 pigments, and the structure for penta-PBS contains 288 protein chains and 348 pigments, whereas OCP has only one chain and one carotenoid.

Among paralogs of OCP including OCP1, OCP2, and OCP3 (14, 15), most OCPs belong to the OCP1 family, which is also the subject of this work (hereafter OCP). In the inactive "orange" form of OCP, called $OCP^O$, a carotenoid spans the N-terminal domain (NTD) and the C-terminal domain (CTD) which are connected by a flexible peptide linker (16). Upon illumination by blue-green light, hydrogen bonds between the CTD and the carotenoid are perturbed, the CTD and NTD separate with the carotenoid burrowing into the NTD (17). The translocation of the carotenoid fully into the NTD causes the changes in the absorption properties of the carotenoid, with the activated form appearing red, known as $OCP^R$. The surrounding protein environment of the carotenoid in OCP$^R$ tunes the transition dipole moment of the carotenoid for quenching (18). The separation of the NTD and CTD also exposes the PBS binding site of OCP in the OCP$^R$ form. OCP$^R$ then binds the phycobilisome (19) and dissipates excess energy as heat. In this work, we will refer to OCP1 from *Synechocystis* sp. PCC 6803 as "tri-OCP" and OCP1 from *Anabaena* sp. PCC 7120 as "penta-OCP". The sequences of the NTD for tri-OCP and penta-OCP are 84% identical (***SI Appendix Fig. S2***) and their OCP$^O$ crystal structures are nearly identical with an RMSD of 0.34 Å (***SI Appendix Fig. S3***).

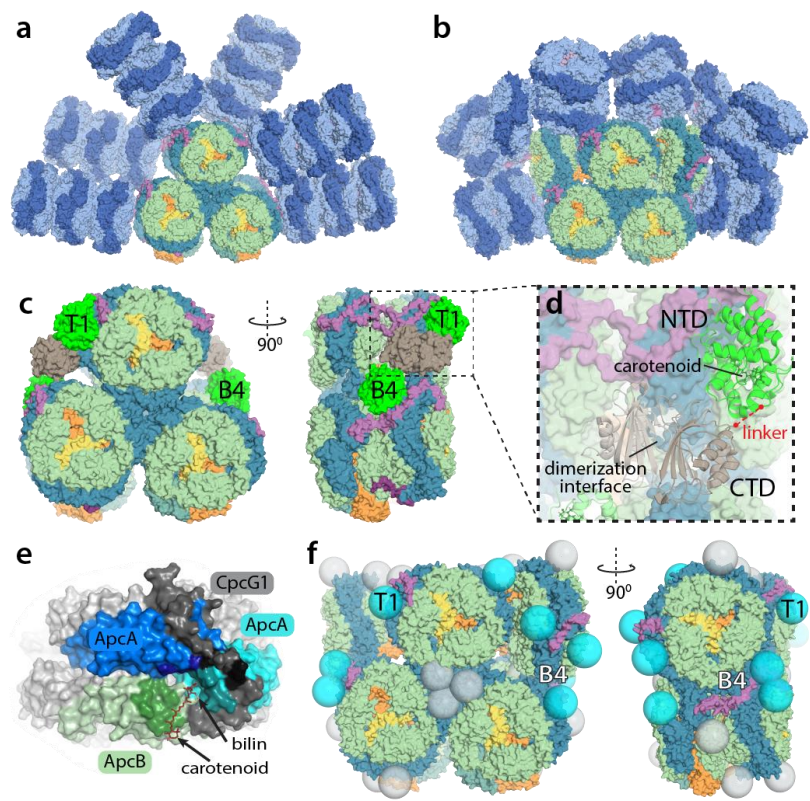

**Figure 1. Structure and OCP binding sites on a tricylindrical and pentacylindrical core phycobilisome.** Cryo-EM structures of **(a)** *Synechocystis* PCC. 6803 phycobilisome (tri-PBS) with a tricylindrical core (composite map from (19)) and **(b)** *Anabaena* PCC. 7120 phycobilisome (penta-PBS) with a pentacylindrical core (PDB: 7EYD (13)) are shown. In addition to the common Apc tricylindrical core, the penta-PBS core has two additional flanking core half-cylinders and 8 Cpc rods. **(c)** Cryo-EM structure of the tri-PBS with 3-OCP$^R$ (lime/tan; PDB: 7SC9 (19)) shows that four tri-OCP$^R$ bind to the tri-PBS core as two dimers, with the NTD of each tri-OCP$^R$ contacting a core cylinder. Dimers are formed through the CTD. The T1 and B4 sites are labeled in black text. **(d)** Zoom on T1 binding site (dotted black box from panel c) showing the tri-OCP$^R$ bound structure (PDB: 7SC9 (19)) with the NTD (lime) and CTD (tan) separated and joined by a linker (red dotted, structure not available). The carotenoid is held in the NTD. The second tri-OCP$^R$ can be seen at lower left, dimerized via the CTD with near-C2 symmetry. **(e)** The consensus binding site of tri-OCP$^R$ on tri-PBS consists mainly of one ApcB (green) and two ApcA subunits (cyan, blue) as well as a peripheral interaction with the CpcG1 rod-core linker (dark gray) (Top cylinder complex from PDB: 8TPJ (18)). Residues within 3 Å of the OCP are colored darker than their corresponding subunit color. The OCP$^R$ (not shown) is located in front with only its carotenoid depicted here as sticks. **(f)** The penta-PBS core (PDB: 7EYD (13)) is shown with all identified instances of the OCP binding motif marked with spheres. Cyan spheres: sites without significant steric clashes. Light gray spheres: sites where OCP access would be blocked by a rod, another part of the core cylinder, or the underlying thylakoid membrane. The T1 site is labeled in black text, and the B4 site (occluded by the E cylinder) is labeled in white text.

Over decades, attempts to locate the OCP$^R$ binding site on the phycobilisome using time-resolved fluorescence spectroscopy (20), site-directed mutagenesis (21), computational modeling (22), and single-molecule fluorescence spectroscopy (23) arrived at the consensus that OCP$^R$ binds to the phycobilisome core. However, until recently the exact binding site(s) and binding stoichiometry proved difficult to pinpoint; the highly symmetric subunit composition of core cylinders means that most candidate binding motifs appear in many locations on the core, yet experimental evidence suggested the presence of only one (6) or two (24) binding locations. Recently, cryo-EM structures of tri-PBS quenched by tri-OCP$^R$ have provided the first definitive insight into the structure of an OCP-phycobilisome complex (4), showing two dimers of tri-OCP$^R$ bound at two pairs of binding sites in the tricylindrical core (**Fig. 1c**). Due to the C2 symmetry of the PBS there are only two unique sites, one on the bottom cylinder, called site "B4", and one on the top cylinder, called site "T1". While the N-terminal domain of the OCP interacts directly with the PBS, the C-terminal is responsible for dimer formation in the space between the PBS core cylinders (**Fig. 1d**). The linker joining the NTD and CTD is flexible and wraps around the NTD before leading to the CTD. The binding interface is almost identical for both unique binding sites and involves two ApcA subunits and one ApcB subunit (**Fig. 1e**). Surprisingly, the bottom cylinder binding site appears to be inaccessible on the penta-PBS, with the flanking core cylinder sterically blocking access. Therefore, even as the discovery of the tri-OCP$^R$ binding sites has advanced our understanding of the quenching mechanism for phycobilisomes with a tricylindrical core, it has also raised a broader question of whether the OCP quenching sites and mechanism are conserved across cyanobacteria with different phycobilisome architectures.

To compare OCP function across species of cyanobacteria, it is necessary to disentangle differences in quenching strength from binding affinity and any underlying photophysical or structural differences among phycobilisomes, which is difficult to achieve through bulk assays. Single-particle fluorescence spectroscopy enables characterization of individual photophysical states without ensemble averaging or synchronization, revealing population-wide or dynamic heterogeneities underpinning molecular function. The handful of single-particle studies performed to date on tricylindrical phycobilisomes and OCP-phycobilisome interactions have discovered photoresponsive dynamics both at the level of individual phycobiliprotein subunits and for the entire phycobilisome, as well as multiple quenched states caused by OCP and evidence of transition states during OCP binding (23–29). Single-particle measurements of the properties and distributions of discrete photophysical states can be compared to computational models based on established molecular structures and bulk ultrafast measurements of energy transfer to understand the molecular mechanisms driving each observed state or transition.

Here, in order to learn whether OCP quenches phycobilisomes by similar mechanisms and at similar sites across cyanobacterial species with different phycobilisome core architectures, we characterized and modeled the photophysical states accessible to prototypical tricylindrical (*Synechocystis* PCC. 6803; tri-PBS) and pentacylindrical phycobilisomes (*Anabaena* PCC. 7120; penta-PBS), with and without OCP, at the single-particle level. To eliminate possible perturbative effects of surfaces or tethers, we use an Anti-Brownian ELectrokinetic (ABEL) trap to measure discrete states in brightness, lifetime, and spectral emission for individually trapped phycobilisomes and phycobilisome-OCP$^R$ complexes. To understand the molecular origin of each state, we compare our experimental results to photon-by-photon Monte Carlo simulations of exciton transfer through compartmentalized models of the phycobilisomes with OCP$^R$ bound at different locations. We found that some features of OCP quenching including dimeric binding and quenching strength are conserved for both architectures, to the point where penta-OCP$^R$ quenches both penta-PBS and tri-PBS with similar efficiency, indicating that its quenching mechanism is likely preserved. However, the dimeric binding site appears to differ, and cross-species quenching is apparently asymmetric since tri-OCP$^R$ is only able to quench tri-PBS.

## Results

### Potential binding locations of OCP on a pentacylindrical core

To assess potential binding sites of the NTD of penta-OCP$^R$ on penta-PBS we located all instances on the penta-PBS core (PDB: 7EYD (13)) of the subunit configuration shown in **Fig. 1e**, in which one ApcB and two ApcA subunits form the previously reported binding site for the NTD of tri-OCP$^R$ on tri-PBS (18, 19). Of the 44 total instances of this motif on penta-PBS, only 7 unique sites (14 total due to the C2 symmetry) have sufficient space to accommodate the OCP$^R$ NTD, two on the bottom (B) cylinder, two on the top (T) cylinder, and three on the extra (E) cylinder (**Fig. 1f;** see ***SI Appendix Fig. S4***).

As previously reported (19), the B4 site, which is equivalent to the bottom cylinder site of the tri-OCP$^R$, is completely blocked by the E cylinder in penta-PBS. However, the top cylinder site for the tri-PBS is the T1 site, which is accessible in penta-PBS. We next evaluated the alignment of subunit chains between the recent high resolution structure of OCP$^R$ bound to the tri-PBS T cylinder (PDB: 8TPJ, (18)) and each accessible binding site on penta-PBS (PDB: 7EYD (13)). This evaluation of potential binding sites assumes similarity of the NTD structures for tri-OCP$^R$ and penta-OCP$^R$ based on their sequence and structural homology. Structural alignment was performed with the consensus binding site consisting of two ApcA subunits and one ApcB subunit. See ***SI Appendix Note S1***, ***SI Appendix Figs. S5-S7***, and ***SI Appendix Table S1*** for a detailed description of evaluation of sites on both tri-PBS and penta-PBS. The penta-PBS T1 site is most similar to the tri-PBS T1 site with an RMSD of 0.68 Å, the other candidate penta-PBS sites align as follows: T3: 0.89 Å, B1: 0.95 Å, B9: 2.25 Å, E2: 1.21 Å, E3: 1.21 Å, and E5: 1.17 Å. Full alignment results are shown in ***SI Appendix Table S2***. We next turned to single-particle experiments and simulations to distinguish whether each of these sites could be plausible candidates for penta-OCP$^R$ binding.

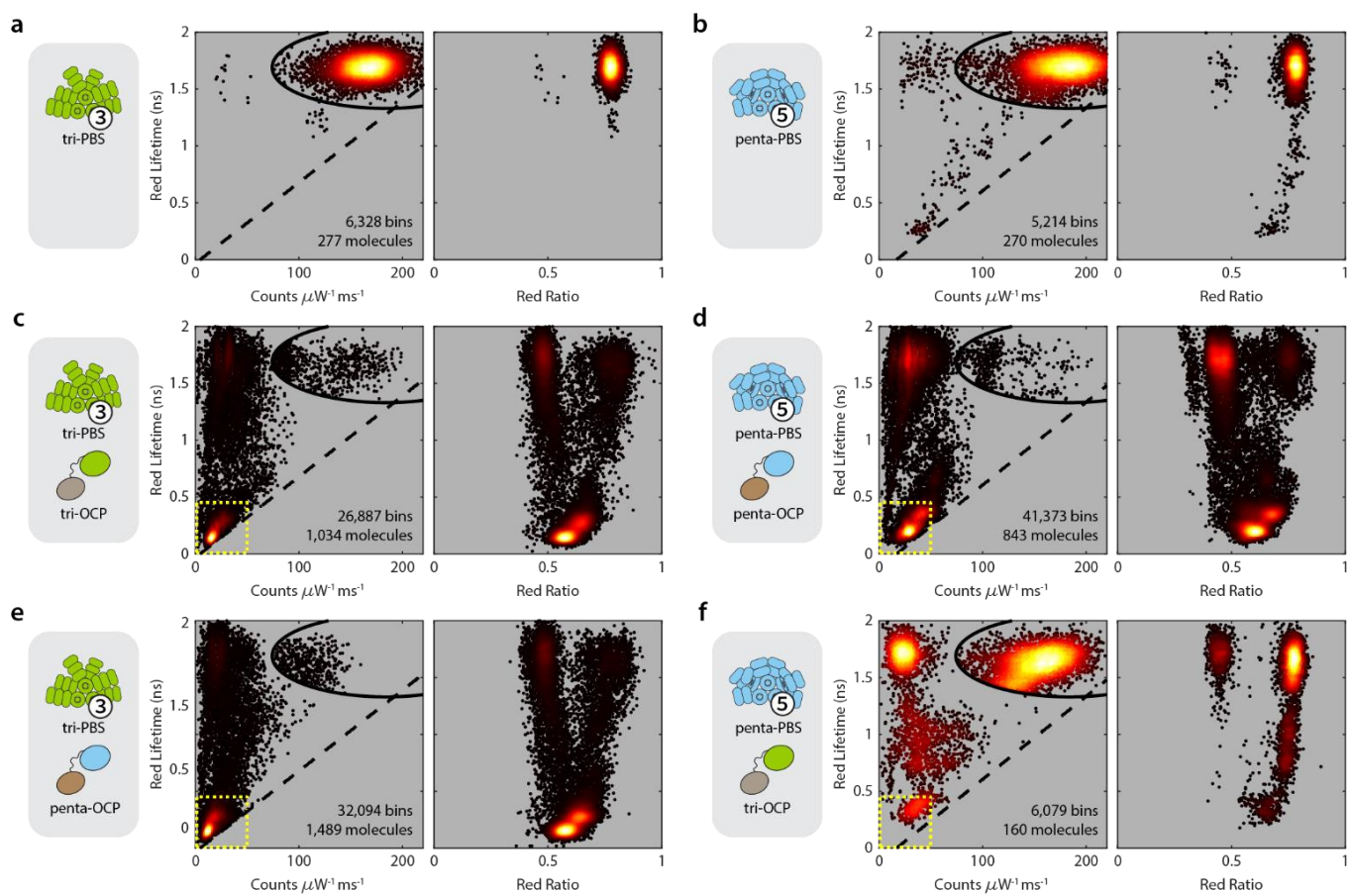

**Figure 2. Scatter heatmaps showing the photophysical states of quenched and unquenched phycobilisomes. a)** Red lifetime-brightness (left) and red lifetime-red ratio (right) projections for tri-PBS are shown as scatter heatmaps, with points colored according to the relative density of neighboring points. Each point represents a 500-photon bin for which the photophysical parameters were calculated. Red lifetime refers to the lifetime calculated from photons detected by the red channel. Data from phycobilisome oligomers was filtered out (black oval; black dotted line; *SI Appendix Figs. S9-S10*; see **Methods** for details). Similarly, the heatmaps for **b)** penta-PBS **c)** tri-PBS quenched by tri-OCP **d)** penta-PBS quenched by penta-OCP **e)** tri-PBS quenched by penta-OCP and **f)** penta-PBS quenched by tri-OCP are shown.

## Measuring single phycobilisomes in an ABEL trap

We used an ABEL trap to observe single phycobilisomes for long timescales without immobilizing or tethering them to a surface, to minimize perturbations that could affect the protein complex structure, energy transfer pathways, or OCP binding (see **Methods** and *SI Appendix Fig. S8*). The ABEL trap uses closed-loop feedback voltages to maintain the position of a single phycobilisome in free solution while monitoring its fluorescence. For each trapped molecule, we measure fluorescence brightness, lifetime, and the ratio of emitted light >660 nm to total brightness, here called the "red ratio". Together, these parameters enable us to distinguish among different types of trapped particles and their photophysical states, including unquenched phycobilisomes with different architectures, various quenched states of the phycobilisome, Cpc rods, and smaller phycobilisome subunits.

In the absence of OCP, both tri-PBS and penta-PBS largely populated the spectroscopic state associated with emission of a single pristine phycobilisome (**Fig. 2a-b**), with some higher order oligomers present that were filtered out during analysis (*SI Appendix Note S2* and *SI Appendix Figs. S9-S10*). Single pristine penta-PBS are approximately 10% brighter than single pristine tri-PBS, averaging 181 ± 26 and 167 ± 20 counts/µW/ms respectively (errors represent standard deviation of gaussian fits). The prototypical tri-PBS structure contains 18 Cpc hexamers across six rods (30, 31, 8). Recent cryo-EM structures for penta-PBS truncate the Cpc rod length to two Cpc hexamers per rod and one Cpc hexamer attached to each of the two Apc half-cylinders due to a combination of rod length heterogeneity, conformational flexibility, and Cpc hexamer dissociation during sample prep (12, 13). Our measurements therefore suggest that our pristine penta-PBS contains 20 Cpc hexamers on average, supporting a model where six of the Cpc rods contains three hexamers each, with the additional two rods formed by one Cpc hexamer each attached to the two extra Apc half-cylinders.

Fluorescence lifetime (1.70 ± 0.10 ns) and red ratio (0.78 ± 0.02) in the pristine state were nearly identical for both types of pristine phycobilisomes (**Table 1**). We attribute small deviations in individual phycobilisome-to-phycobilisome brightness to variability of rod lengths. Notably, raw data traces reveal that each phycobilisome displays consistent photophysical parameters while trapped, confirming that the ABEL trap does not perturb the structure of trapped phycobilisomes and that the low excitation laser power used here does not induce photophysical dynamics or damage (*SI Appendix Figs. S11-S12*).

Table 1. Mean and standard deviations of measured parameters for unquenched, quenched, and rod populations from experiments shown in **Fig. 2**.

| Phycobilisome | OCP | State | Brightness (counts ms$^{-1}$ µW$^{-1}$) | τ Red channel (ns) | τ Green channel (ns) | Red ratio |
|---|---|---|---|---|---|---|
| *Synechocystis* PCC. 6803 (tricylindrical; tri-PBS) | No OCP | Unquenched | 167 ± 20 | 1.70 ± 0.09 | 1.32 ± 0.11 | 0.775 ± 0.023 |
| | *Synechocystis* PCC. 6803 (tri-OCP) | Q1 | 25.0 ± 5.1 (15%) | 0.28 ± 0.04 | 0.20 ± 0.03 | 0.645 ± 0.032 |
| | | Q2 | 15.9 ± 3.0 (9.5%) | 0.15 ± 0.03 | 0.11 ± 0.03 | 0.577 ± 0.032 |
| | | Rods | 29.0 ± 7.2 | 1.72 ± 0.14 | 1.67 ± 0.12 | 0.482 ± 0.026 |
| | *Anabaena* PCC. 7120 (penta-OCP) | Q1 | 19.4 ± 4.9 (12%) | 0.28 ± 0.04 | 0.21 ± 0.03 | 0.640 ± 0.034 |
| | | Q2 | 12.23 ± 3.0 (7%) | 0.16 ± 0.03 | 0.11 ± 0.03 | 0.578 ± 0.034 |
| | | Rods | 20.6 ± 6.3 | 1.68 ± 0.15 | 1.61 ± 0.15 | 0.486 ± 0.027 |
| *Anabaena* PCC. 7120 (pentacylindrical; penta-PBS) | No OCP | Unquenched | 181 ± 26 | 1.70 ± 0.10 | 1.34 ± 0.13 | 0.783 ± 0.023 |
| | *Anabaena* PCC. 7120 (penta-OCP) | Q1 | 40.5 ± 6.6 (22%) | 0.35 ± 0.04 | 0.24 ± 0.03 | 0.673 ± 0.034 |
| | | Q2 | 28.4 ± 5.6 (16%) | 0.21 ± 0.04 | 0.14 ± 0.03 | 0.597 ± 0.036 |
| | | Rods | 29.3 ± 7.9 | 1.71 ± 0.11 | 1.66 ± 0.11 | 0.453 ± 0.036 |
| | *Synechocystis* PCC. 6803 (tri-OCP) | Unquenched | 159 ± 24 | 1.66 ± 0.10 | 1.47 ± 0.11 | 0.775 ± 0.022 |
| | | Q1 (possible) | 36.6 ± 9.0 (21%) | 0.36 ± 0.08 | 0.25 ± 0.04 | 0.656 ± 0.043 |
| | | Rods | 24.2 ± 6.3 | 1.72 ± 0.11 | 1.68 ± 0.11 | 0.447 ± 0.030 |

**Tri-PBS and penta-PBS exhibit similar OCP-quenched photophysical states**

To establish the reference states for OCP quenching in a tricylindrical phycobilisome, we first examined single tri-PBS in the presence of activated tri-OCP, illuminated at high laser power in the ABEL trap. At these higher laser powers, non-quenched phycobilisomes exhibit photodynamics as well as photobleaching over time due to photodamage and can exhibit states that overlap with OCP-quenched states (*SI Appendix Figs. S13-S14*), so the photobleached portions of these events are filtered out during analysis for all OCP quenching experiments (*SI Appendix Note S2* and *SI Appendix Figs. S15-S17*). We observed that tri-PBS was quenched as expected (**Table 1**; **Fig. 2c**), with three distinct photophysical states emerging: a dim, long-lifetime, and highly blue-shifted state, B, likely corresponding to detached Cpc rods, and two quenched populations which we label Q1 (15% brightness compared to unquenched phycobilisome) and Q2 (9.5% brightness). Here, Q1 and Q2 are brighter than for the previously reported "truncated" *Synechocystis* PCC. 6803 phycobilisome which only contains one Cpc hexamer per rod (11% and 6% brightness, respectively) (24), in qualitative agreement with previous reports using bulk quenching assays for both phycobilisome variants (6). The lifetimes of Q1 and Q2 are short, and the red ratio is blue-shifted compared to the pristine state, as expected. In the context of the known structure of the tricylindrical phycobilisome-OCP complex (14), and supported by our computational model (*vide infra*), we interpret Q1 as a state where one OCP dimer is bound at the established binding location, and Q2 as the fully-quenched state with two OCP dimers bound.

Because the tri-OCP binding sites are sterically blocked in the penta-PBS cryo-EM structure, we expected that a single penta-PBS observed under identical excitation conditions in the presence of its own activated penta-OCP might exhibit shifted, or different numbers of, quenched states. However, the quenched penta-PBS also produced three photophysical states that appeared to correspond directly to the B, Q1, and Q2 states that were observed for the quenched tri-PBS (**Table 1**; **Fig. 2d**). For penta-PBS, the Q1 state (22% brightness compared to unquenched penta-PBS) and Q2 state (16% brightness) are proportionally less quenched than in the tri-PBS. This shift could be caused by differences in the quenching site(s) or molecular mechanism, or by differences in the architecture and energy transfer rates within the two phycobilisome species. The penta-PBS has more Cpc hexamers and slower rod-to-core exciton transfer rate than the tri-PBS (32), which could produce a higher proportion of rod fluorescence in the quenched penta-PBS. The lifetimes and red ratio measured for the penta-PBS Q1 and Q2 are nearly identical to tri-PBS Q1 and Q2 photophysical states.

**Tri-PBS are near-identically quenched by both the tri-OCP and penta-OCP**

Since both the phycobilisome and OCP are highly modular, and cross-species quenching is known to occur (15, 33, 34), we next tested quenching of tri-PBS by activated penta-OCP to see how the resulting quenched states, if any, might differ from quenching by the native OCP. This combination also produced two states closely corresponding to Q1, and Q2 (**Table 1**; **Fig. 2e**). Both Q1 and Q2 are slightly dimmer for the penta-OCP (12% and 7% brightness, respectively) than for the tri-OCP, but the lifetimes and red ratios are nearly identical. These data support that the penta-OCP likely interacts with the tri-PBS at the same sites as the tri-OCP and quenches in a similar manner, which is consistent with the high sequence homology between the two OCPs. However, an alternative possibility remains that penta-OCP quenches tri-PBS by a different combination of binding sites and quenching rate(s) that happen to produce the same number and photophysical properties of quenched states as the native tri-OCP.

**Penta-PBS are not quenched by tri-OCP**

Notably, we did not observe quenched populations when activated tri-OCP was added to the penta-PBS (**Fig. 2f**; **Table 1**). This result is surprising because the tri-PBS was fully quenched by both types of OCP, which might otherwise suggest mutual compatibility of the phycobilisome-OCP systems. Instead, the observed one-way compatibility indicates that even though penta-OCP can access, bind, and quench the tri-PBS, the tri-OCP either cannot bind to the penta-PBS quenching sites, or does not have access to them, or can access and bind but does not quench.

**Q1 and Q2 correspond to differing numbers of bound $OCP^R$**

To verify that Q1 and Q2 are produced by different numbers of bound $OCP^R$ rather than by different quenching sites, we titrated the penta-$OCP^R$ with both the penta-PBS and tri-PBS (***SI Appendix Fig. S18***). For both types of phycobilisome in the presence of excess activated OCP, most phycobilisome are in the Q2 state. As the OCP concentration is reduced, the quenched population shifts to favor the Q1 state. Although the penta-OCP fully quenches both types of phycobilisomes, it appears to have higher affinity for its native penta-PBS than for the tri-PBS (79% Q2 vs. 65% Q2 at ~1:100 phycobilisome:OCP). Together, these data confirm that Q1 and Q2 are likely produced by increasing numbers of $OCP^R$ bound to the phycobilisome.

**The isolated NTD of OCP (RCP) produces additional, less-quenched states**

For tri-PBS, Q1 and Q2 likely represent binding of one or two tri-OCP dimers, respectively; if a monomer could bind and quench the phycobilisome then presumably at least one even less-quenched state would be observed at low OCP concentration. For penta-PBS, the oligomeric state of bound OCP is not known. Previous work has shown that the N- and C-terminal domains of OCP can be cleaved using a protease, and that the resulting N-terminal domain, called red carotenoid protein (RCP), can quench the phycobilisome (35). Without the C-terminal domain (CTD) the RCP is constitutively active and the potential for dimerization is lost. Therefore, to test if the penta-OCP binds as a dimer or monomer, we next performed single-molecule quenching assays using cleaved OCP. ***SI Appendix Figs. S19-S20*** show unquenched and quenched bulk spectra for all four phycobilisome-OCP and phycobilisome-RCP combinations. Note that unlike OCP, RCP unbinds from the phycobilisome at the low concentrations needed for single-molecule measurements (***SI Appendix Fig. S21***), so additional RCP was added to the sample during single-molecule measurements in the ABEL trap.

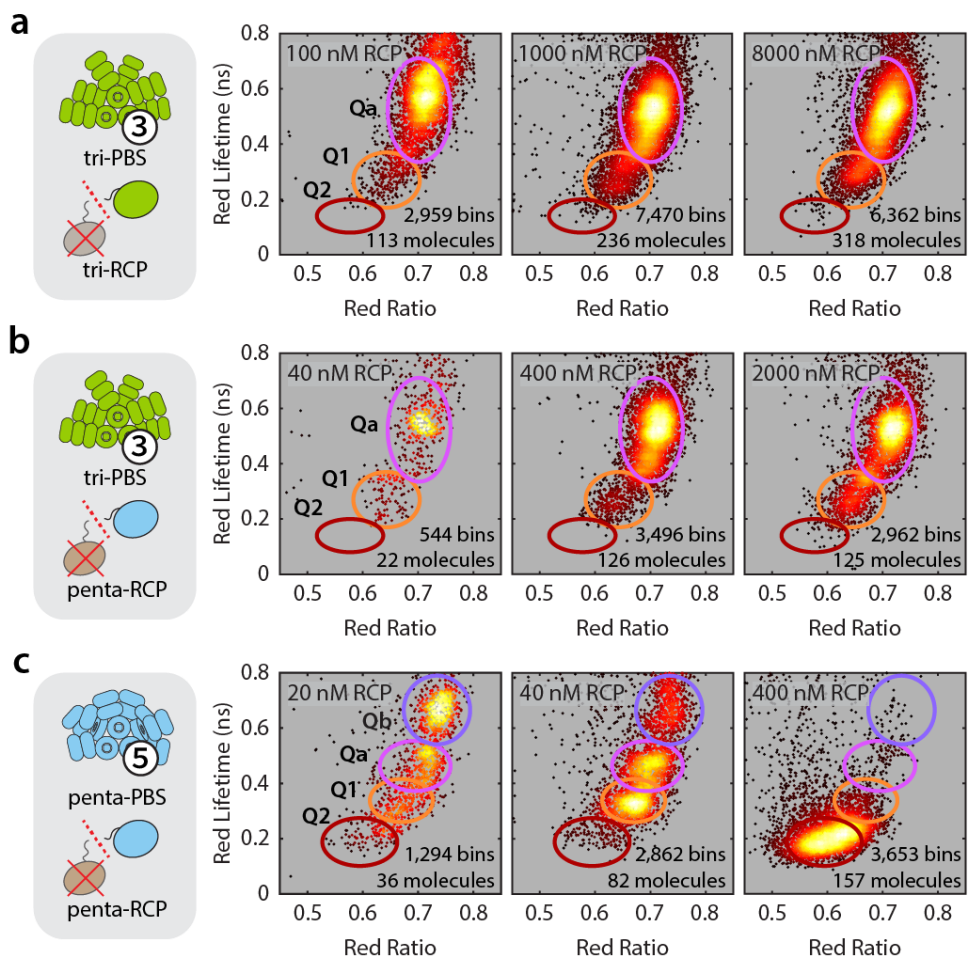

**Figure 3. Titrations with the permanently active red carotenoid protein (RCP). a)** The tri-PBS is not fully quenched with the tri-RCP. Even at extremely high concentrations of tri-RCP, the Q2 (red ellipse) state is not observed and relatively few phycobilisomes visit the Q1 (orange ellipse) state. Most phycobilisomes remain in a newly emerged less quenched state, labeled Qa (pink ellipse). **b)** Similar results were observed upon quenching the tri-PBS with the penta-RCP. **c)** On the other hand, when the penta-RCP is present in excess, penta-PBS can be fully quenched to the Q2 state. As the amount of RCP is reduced, the population shifts to lower quenched states including the two new states, Qa and Qb (purple ellipse). Additionally, unlike OCP, RCP dissociates quickly from the phycobilisome at the low concentrations needed for single-molecule studies, thus requiring the presence of excess RCP during measurement (***SI Appendix Fig. S21***).

When the tri-PBS is quenched by *Synechocystis* PCC. 6803 RCP (tri-RCP), we observe a new, less quenched state, labeled Qa (**Fig. 3a**). Surprisingly, even with a large excess of RCP, we observe neither Q1 nor Q2 when quenching with tri-RCP. At high tri-RCP concentrations, a slight tail on Qa encroaches towards Q1, but it is not clear whether or not this population represents a separate quenched state. This result suggests that in the tri-PBS, the dimerization of OCP through the CTD is crucial either for accessing some of the binding sites, or for quenching, or both. Tri-PBS with penta-RCP also exhibited only population Qa regardless of RCP concentration (**Fig. 3b**), suggesting that removal of the penta-OCP's CTD inhibits binding and/or quenching in a similar manner to tri-OCP, and further supporting possible mechanistic similarities between quenching by the NTD of tri-OCP and of penta-OCP. As with tri-OCP, the tri-RCP did not quench the penta-PBS at all. This asymmetry suggests that the tri-NTD cannot interact with the penta-PBS in the same way as the NTD of penta-OCP[R].

Titrating penta-PBS with *Anabaena* PCC. 7120 RCP (penta-RCP) revealed at least four distinct quenched populations (**Fig. 3c**). At the highest concentrations of RCP, penta-PBS was quenched to the Q2 photophysical state that was originally observed using the full OCP. Lowering the amount of RCP generated the Q1 state as well as two additional less-quenched populations, Qa and Qb, where Qa is more quenched than Qb. Since the Qa and Qb states are only observed when quenching with the RCP, which is an obligatory monomer, and because Qa and Qb are less quenched than Q1 and Q2, we conclude that these additional states are likely produced by single RCPs binding and quenching the phycobilisome. The appearance of two new populations suggests that monomers or RCPs may bind at more than one site, thereby effecting different quenching levels.

Table 2. Mean and standard deviations of extracted parameters of observed quenched states from experiments shown in **Fig. 3**.

| Phycobilisome | RCP | State | Br (counts ms$^{-1}$ μW$^{-1}$) | $\tau$ Red channel (ns) | $\tau$ Green channel (ns) | Red ratio |
|---|---|---|---|---|---|---|
| *Synechocystis* PCC. 6803 (tricylindrical; tri-PBS) | *Synechocystis* PCC. 6803 (tri-RCP) | Qa | 41.3 ± 9.5 (25%) | 0.524 ± 0.085 | 0.392 ± 0.092 | 0.708 ± 0.023 |
| | *Anabaena* PCC. 7120 (penta-RCP) | Qa | 45.6 ± 9.7 (27%) | 0.520 ± 0.079 | 0.391 ± 0.087 | 0.709 ± 0.022 |
| *Anabaena* PCC. 7120 (pentacylindrical; penta-PBS) | *Anabaena* PCC. 7120 (penta-RCP) | Qb | 80.6 ± 11 (45%) | 0.661 ± 0.05 | 0.511 ± 0.07 | 0.737 ± 0.020 |
| | | Qa | 59.4 ± 8.4 (33%) | 0.457 ± 0.04 | 0.330 ± 0.06 | 0.697 ± 0.025 |
| | | Q1 | 46.5 ±7.4 (26%) | 0.336 ± 0.04 | 0.248 ± 0.05 | 0.670 ± 0.024 |
| | | Q2 | 28.3 ± 5.4 (16%) | 0.195 ± 0.03 | 0.148 ± 0.04 | 0.598 ± 0.031 |

## Monte Carlo simulations: Connecting phycobilisome structure to OCP binding and quenching function

The experimental results presented above invite multiple plausible mechanistic interpretations. To facilitate quantitative evaluation of various hypotheses, we developed and benchmarked computational models of both the tri-PBS and penta-PBS (***SI Appendix Note S3*** and ***SI Appendix Figs. S22 and S23***). Following previous modeling efforts (24, 36), we elected to use a compartmental Monte Carlo simulation of energy transfer (**Fig. 4a-b**) where equilibration within each compartment is assumed to be very fast relative to energy transfer between compartments, and rate constants are derived from ultrafast measurements (32, 36). We used these models to predict how different OCP binding sites, phycobilisome-OCP stoichiometry, and OCP quenching rates would influence the experimentally observed quenched states.

We simulated the tri-PBS with four OCPs arranged as two dimers, where one OCP from each dimer quenched a top core cylinder compartment and one OCP from each dimer quenched a bottom cylinder compartment, in agreement with the cryo-EM structures (19). Using the pristine state as our benchmark, we simulated a range of possible OCP quenching rates under the assumption that all four OCPs have the same quenching rate and determined that a quenching rate of 31 ns$^{-1}$ would be necessary to produce the experimentally observed properties of state Q2 (**Fig. 4c**). We verified that this quenching rate also correctly produces state Q1 when only one OCP dimer is attached to the core and state Qa when only one OCP monomer is attached to the core and used this quenching rate for subsequent simulations.

Other combinations of two and four binding sites on tri-PBS produce states that are remarkably similar to Q1 and Q2 (**Fig. 4c** and ***SI Appendix Fig. S24***), respectively, with the notable exception of the BB' (two OCPs bound, top cylinder only) and AC (two bound, bottom of one phycobilisome face only), which produce a state that is less quenched than Q1 but more quenched than Qa. When only one OCP is present, state Qa is predicted regardless of which core compartment is quenched. Together, the results of these tri-PBS simulations establish an approximate quenching rate for the tri-OCP monomer, validate the interpretation of Q1 and Q2 as one and two bound OCP dimers, respectively, and support the hypothesis that state Qa is indeed generated by a single RCP bound at an unknown location on the core.

The quenching simulations for tri-PBS also provide insight into our results for penta-PBS quenching experiments. Since an identical Qa state was observed for penta-RCP binding to tri-PBS, and since simulated quenching at any core compartment produced Qa for an OCP monomer, penta-OCP likely quenches with a similar rate to tri-OCP. Although our simulations showed that Q1 and Q2 cannot be used to differentiate among combinations of 2- and 4 binding sites, experimental observation of Q1 and Q2 for penta-OCP on tri-PBS establishes that it likely binds as a dimer.

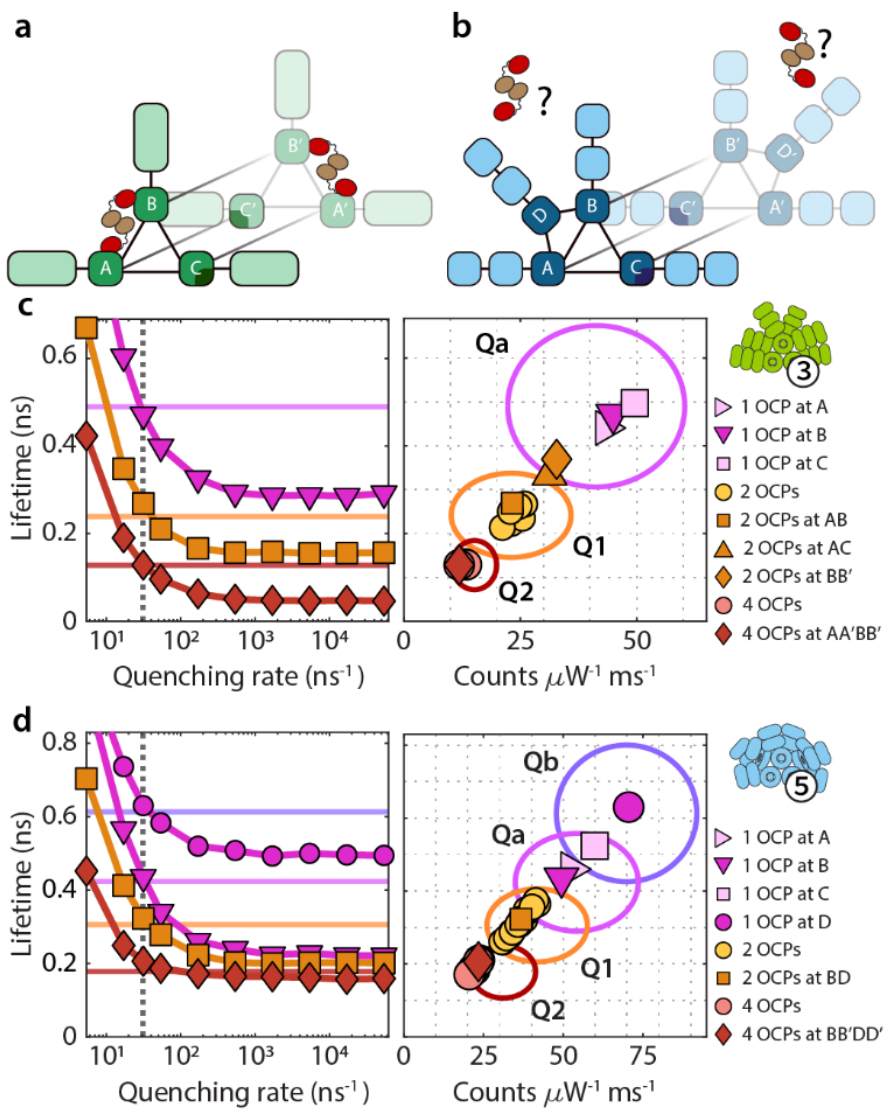

**Figure 4. Photon-by-photon Monte Carlo simulations.** A compartmental model for **(a)** tri-PBS and **(b)** penta-PBS was used to simulate the flow of photons through the phycobilisomes using experimentally determined transition rates (32, 36) with and without OCP(s) attached to the different compartments. **c)** For the tri-PBS, the known binding sites and OCP/phycobilisome stoichiometry were used to simulate the quenched phycobilisomes with two OCP dimers with varying quenching rates. Vertical dotted line indicates the quenching rate (31 ns$^{-1}$) where the simulated lifetime for tri-PBS quenched by 2 OCP dimers intersects the experimentally determined lifetime for Q2 state. At a quenching rate of 31 ns$^{-1}$, one OCP bound to a bottom or the top cylinder could produce the experimentally observed Qa state. Similarly, one OCP dimer bound to one of the bottom and top cylinders could produce the experimentally observed Q1 state. **d)** For the penta-PBS, one, two, or four OCPs were attached to different compartments or combinations of compartments. Assuming the penta-OCP has the same quenching rate as the tri-OCP (dotted line), four OCPs are needed to reach the experimentally determined lifetime for the Q2 state. At a quenching rate of 31 ns$^{-1}$, one OCP bound to the flanking cylinders could produce the Qb state while one OCP on the top cylinder could produce the Qa state. The Q1 and Q2 states likely correspond to one and two dimers bound to the penta-PBS core, respectively.

Although the quenching of penta-OCP may not be identical on tri-PBS compared to penta-PBS, the rates are likely to be similar. Therefore, we next simulated penta-OCP on penta-PBS across a range of rates, to test whether our simulations matched experimental data at a similar quenching strength. We simulated one, two, or four OCPs at all combinations of phycobilisome compartments (**Fig. 4d** and *SI Appendix Fig. S25*). Here, too, a quenching rate of 31 ns$^{-1}$ generated states that closely matched our experimental results. Most importantly, we found that one OCP at the flanking core half-cylinders compartments D or D' produced the Qb state, while one OCP on any of the top (B) or bottom (A or C) compartments produced the Qa state. Our experimental observation of both Qa and Qb for the penta-PBS quenched by penta-RCP, then, suggests that compartments D and D' must be viable binding sites on this phycobilisome. Simulating two or four OCPs consistently produced the Q1 and Q2 states, respectively, and as with tri-PBS, the combinations of quenched compartments could not be well-differentiated. We note that at a higher quenching rate (about 3 times higher), two OCPs could in principle achieve the Q2 state, and one OCP the Q1 state, but in this case the Qb state would remain unexplained.

### Evaluation of dimeric binding sites on penta-PBS

Finally, we evaluated all possible combinations of the seven available sites to determine which, if any, pairs might be good candidates for binding dimers of OCP (*SI Appendix Table S3-S4*). The dimeric binding site on tri-PBS involves sites T1 and B4 (B4 is inaccessible on penta-PBS), with a distance of 84.8 Å between the NTD centers of mass. Of all 21 pairs evaluated for penta-PBS, most can be eliminated on the basis of their NTD spacing (49 Å <

spacing < 97 Å) or because simulations of quenching at that pair of sites are inconsistent with the observed quenched states (for example, pairs that quench the same compartment). The most likely candidates after applying these filters are pairs T1 and B1 (**Fig. 5a**); T1 and E3 (**Fig. 5b**); as well as the three pairs shown in *SI Appendix Fig. S26*: T3 and E5'; B9 and E2; and B9 and E5. However, in the tri-OCP$^R$ bound structure, the linkers connecting the NTDs and CTDs wrap around the NTDs (**Fig. 1d**). If that portion of the linker was instead fully extended, the distance between NTDs would increase, and the global conformation of each monomer would change, enabling access to other binding site combinations for the OCP$^R$ dimer that are not considered here.

Of these five pairs of sites, we propose that the two combinations involving T1 seem most likely since the T1 site is used in tri-PBS binding with tri-OCP (see *SI Appendix Note S4* for discussion of binding sites involving T3 and B9). As noted above, a crucial difference between these two possibilities is the orientation of the linkers connecting the NTD and the CTD of the two OCPs. In the tri-PBS structure, the OCP linkers of the bound OCPs point towards each other with a relative linker orientation of 146° (*SI Appendix Table S3*). As indicated by the gray vectors in **Fig. 5**, binding of an OCP dimer to T1 and B1 would place the NTD to CTD linkers in a parallel orientation (8° relative orientation), eliminating the dimer's C2 symmetry (*SI Appendix Table S4*). By contrast, the relative linker vectors for the T1 and E3 sites are nearly anti-parallel (153°), similar to tri-PBS.

Moreover, our titration experiment with RCP showed additional quenched populations, Qa and Qb, and simulations revealed that the Qb population is likely the result of an RCP bound to the E cylinder. Thus, assuming RCP binds to the same sites as OCP, one of the OCP-binding sites must be on the E cylinder, making T1 and E3 the most likely binding site pair for an OCP dimer.

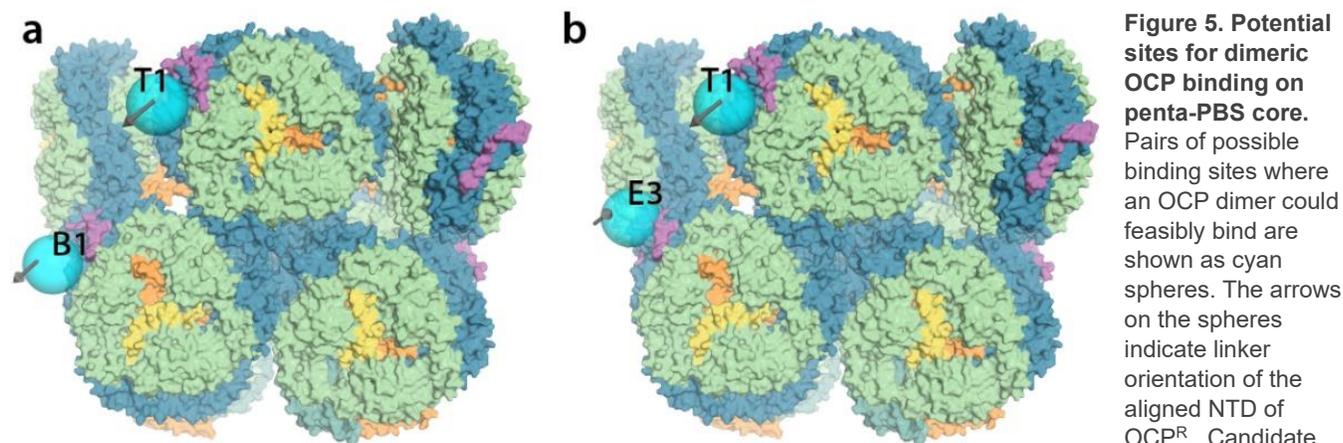

Figure 5. Potential sites for dimeric OCP binding on penta-PBS core. Pairs of possible binding sites where an OCP dimer could feasibly bind are shown as cyan spheres. The arrows on the spheres indicate linker orientation of the aligned NTD of OCP$^R$. Candidate binding sites were identified by aligning the known binding site on the tri-PBS with the same binding motif (two ApcA and one ApcB) on the penta-PBS. **(a)** Sites T1 and B1 (NTD spacing: 96.7 Å; linker orientation: 8°). **(b)** Sites T1 and E3 (NTD spacing: 79.5 Å; linker orientation: 153°).

## Discussion

Here we have combined insights from single-particle characterization, numerical modeling, and structural biology data to understand the photophysical states—and to deduce the sites—accessible to penta-PBS and tri-PBS in the presence of penta-OCP and tri-OCP. We found that OCP appears to quench with similar strength and in similar numbers on both phycobilisome architectures, but must quench at different locations on the phycobilisome core. Although detailed structural studies will be necessary to confirm the exact binding site(s) of penta-OCP on penta-PBS, our work highlights the ability of a combined bottom-up approach to reveal mechanistic details of dynamic and heterogeneous molecular interactions.

One major remaining puzzle is the inability of the tri-OCP and tri-RCP to quench the penta-PBS. Especially given that both the OCP sequence and residues in the canonical binding site are highly conserved and we observed that penta-OCP is able to quench tri-PBS almost indistinguishably from its native tri-OCP. Additionally, the accessible OCP-binding site on the top cylinder of penta-PBS is only slightly different (0.68 Å RMSD) from the closest related known OCP-binding site on tri-PBS. One possibility is that a tri-OCP dimer cannot adopt a conformation appropriate to the penta-PBS binding site. Alternatively, since our assumptions rely only on the structure of the tri-OCP bound tri-PBS complex, it is very possible that the penta-OCP bound penta-PBS complex is slightly different in a way that would prevent the tri-OCP from binding.

Overall, our results show that the mechanism and strength of OCP quenching appears to be conserved across species of cyanobacteria with dissimilar phycobilisome core architectures, and that the binding site at which quenching occurs appears to be partially conserved. This work lays a foundation for future study of the potentially complex interplay of phycobilisomes with the many OCP homologs that have been identified, which may quench at alternative sites or with different rates than the OCP1 variants studied here, and with other accessory proteins such as the fluorescence recovery protein, FRP. These results underscore the exquisite modularity balanced with adaptability of the phycobilisome, and the corresponding specificity and efficiency of OCP quenching. Phycobilisome-OCP quenching is a unique system that could inspire future artificial light-harvesting technologies to incorporate a wide range of modulation effected by providing a small "fuse" at just the right location.

## Materials and Methods

### Phycobilisome purification

Previously established protocols (6, 37) were used to isolate phycobilisomes from *Synechocystis* sp. PCC 6803 and *Anabaena* sp. PCC 7120. Briefly, cyanobacteria were grown photoautotrophically in BG11 medium at 30 °C and with 3% $CO_2$. The cells were collected by centrifugation (8,000 rpm; 10 min) and washed with 0.8 M phosphate buffer, pH 7.5. After centrifugation, the pellet was resuspended in 0.8 M phosphate buffer containing protease inhibitor and 10 units/ml of DNase I. The cells were broken by passing through a French press twice at 1000 psi. Broken cells were incubated with 1% v/v Triton X for 15 minutes while shaking under dark. Cellular debris was removed by centrifuging at 30,000g for 15 minutes. The supernatant was collected and centrifuged again at 42,000 rpm for 1 hour. The supernatant was added to a discontinuous sucrose gradient containing layers of 1.5, 1.0, 0.75, 0.5, and 0.25 M sucrose in 0.75 M phosphate buffer and centrifuged overnight at 25,000 rpm in a swinging bucket rotor. Intact phycobilisomes were collect at the interface of 0.75 and 1.0 M sucrose layer.

### OCP purification and activation

The OCPs from *Synechocystis* sp. PCC 6803 and *Anabaena* sp. PCC 7120 were expressed in *E. coli* as described previously (19). Briefly, the sequence for a C-terminal histidine tag and the *Synechocystis* sp. PCC 6803 *ocp* gene (*slr1963*) were cloned into pCDFDuet (Novagen). Similarly, the *Anabaena* sp. PCC 7120 *ocp* gene (*all3149*) along with a C-terminal histidine tag were cloned into pET28a (Novagen). The resulting plasmids were expressed in BL21 (DE3) along with the plasmid pAC-CANthipi (Addgene (38)) to obtain the canthaxanthin-containing OCPs. The proteins were purified via affinity chromatography (HisTrap HP, GE Healthcare) and hydrophobic interaction chromatography (HiTrap Phenyl HP, GE Healthcare) to yield the holo-OCP. Purified OCPs were activated using a 488 nm laser (Coherent OBIS 1220123; 10 min; ~1 µmol photons $s^{-1}$ $cm^{-2}$).

### Converting OCP into the permanently active RCP

We cleaved the linker connecting the N-terminal domain (NTD) and the C-terminal domain (CTD) of the OCP to create the permanently active RCP. To cleave the OCP linker, we used trypsin digestion (Promega V5111) following the method from Leverenz et al (35). Briefly, trypsin was added to activated OCP with a protease:protein ratio of 1:150. Protein digestion was allowed to proceed for 20 min under the 488 nm laser illumination. Then, phenylmethylsulfonyl fluoride was added to a final concentration to 1 mM to quench the reaction.

### Phycobilisome quenching assay

$OCP^R$ or RCP was mixed with purified phycobilisomes in excess. The mixture was incubated on ice (in dark) for 10 minutes and diluted to ~ 25-100 pM prior to trapping. When titrating with $OCP^R$ (***SI Appendix Fig. S18***), we reduced the relative amount of $OCP^R$ during incubation. For RCP titration experiments (**Fig. 3**), excess RCP was present during incubation. Since RCP unbinds at the low concentrations needed for single-molecule measurements (***SI Appendix Fig. S21***), varying amounts of RCP (as shown in **Fig. 3**) were present in the sample during measurement. A 1 M phosphate buffer (pH 7.4) with 1 M sucrose was used for binding and single-molecule measurements.

## Single-molecule measurements

The ABEL trap was implemented similar to previous iterations (26, 27). Briefly, an acousto-optic tunable filter (Leukos TANGO VIS) paired with a pulsed supercontinuum laser (Leukos ROCK 400-4) and a pulse picker (Conoptics M350-160-01 KD*P EOM) were used to select 594 nm at 30.1 MHz repetition rate as the excitation source. The excitation laser was passed through two orthogonally placed AODs (MTT110-B50A1.5-VIS) and relayed to the back of the microscope and reflected by a dichroic (Di03-R594-t3-25x36) towards the objective (Olympus UPLSAPO100XS). The fluorescence from a trapped molecule passed through the microscope dichroic, a 300 µm pinhole and the emission filters (Semrock FF01-650/150-25 and NF03-594E-25) and was split using a dichroic (T660lpxr) into red and green channels and detected by two APDs (Excelitas SPCM-AQRH-14-ND). Upon detecting a photon, the APDs send a signal to the TCSPC (Picoquant Multiharp 150) and an FPGA (National Instruments PCIe-78656) to provide feedback. The FPGA calculates the required voltages in the X and Y directions and sends them to two voltage amplifiers (Pendulum F10AD) which apply the voltage to the sample through four platinum electrodes placed orthogonally in a microfluidic chip.

## Microfluidic cell preparation

The microfluidic cells were prepared in-house as previously described (39). Routine photolithography techniques were used to etch the trapping region with ~700 nm depth and surrounding channels with a depth of 12 µm (*SI Appendix Fig. S8*). The microfluidic cells were permanently bonded to quartz coverslips using 6% sodium silicate solution (SIGMA 338443) following procedure adapted from (40). The cells were cleaned and reused indefinitely.

Prior to each trapping experiment, the microfluidic cells were cleaned in piranha (3:1 mixture of sulfuric acid and hydrogen peroxide) overnight. Then, the cells were incubated in 1 M KOH for 10 minutes and rinsed thoroughly. To prevent protein from sticking to the cell walls, we passivated the cells using polyelectrolyte multilayers. First, the cells were incubated in 0.5 wt. % poly(ethyleneimine) solution (Sigma Aldrich 181978) for 10 minutes. The cells were rinsed twice and incubated in 0.5 wt. % poly(acrylic acid, sodium salt) solution (Sigma Aldrich 416010) for 10 minutes. These two steps were repeated one more time. The cells were rinsed with ultrapure water before adding the sample for trapping.

## Data Analysis

All data analysis was performed in MATLAB. The time-tagged photon data from the TCSPC was used to construct 10-ms binned brightness traces. An established change-point finding algorithm (41) was used to find levels of constant brightness within each trace. Each level was divided into 500-photon groups to calculate brightness, lifetime for both channels (42, 43), and red ratio. The red ratio is defined as:

$$\text{Red ratio} = \frac{I_{red}}{I_{red} + I_{green}}$$

where $I_{red}$ is the background subtracted intensity of the red channel (emission photons over 660 nm) and $I_{green}$ is the background subtracted intensity of the green channel (emission photons below 660 nm). All reported brightness values were normalized by dividing by the measured excitation intensity before the microscope objective. Lifetimes for single-molecule data were fitted using single exponential fits convolved with the instrument response function (IRF) determined using a low fluorescence lifetime dye, malachite green. For each 500-photon group, pairs of calculated photophysical parameters were plotted as 2-D scatter heatmaps where each point is colored according to the relative density of neighboring points, as shown in **Figs. 2-3**. Additional details of protein sequence analysis and data analysis are provided in *SI Appendix Note S1-S2*.

## Energy transfer simulations

A detailed description of the Monte Carlo simulations of energy transfer in unquenched and quenched phycobilisomes is provided in *SI Appendix Note S3*.

# Acknowledgements

A.H.S. acknowledges support from the Early Career Award from the Office of Science of the US Department of Energy (DE-SC0025385), as well as support from NSF QLCI QuBBE grant OMA-2121044 and from the Neubauer Family Foundation. C.A.K. acknowledges research support from the Office of Science of the US Department of Energy (DE-SC0020606).

# Author Contributions

A.H.S., C.A.K., and A.E. designed research; A.H.S. and C.A.K. supervised research; A.E. and S.L-Y. generated phycobilisome and OCP samples; A.E. performed experiments; A.E. and M.S. performed structural calculations, A.E. performed Monte Carlo simulations; A.E., M.S., and A.H.S. analyzed data; A.E. and A.H.S. drafted the manuscript; all authors contributed comments and revisions to the manuscript.

# Competing Interest Statement

The authors declare no competing interests.

# Data Availability

Raw data for all figures in the manuscript will be made publicly available on the Zenodo data repository upon publication.

**Supporting Information for**

**Phycobilisome core architecture influences photoprotective quenching by the Orange Carotenoid Protein**

Ayesha Ejaz[1], Markus Sutter[2,3,4], Sigal Lechno-Yossef[2], Cheryl A. Kerfeld[2,3,4,5], Allison Squires[6,7,8]*

[1] Department of Chemistry, University of Chicago, Chicago, IL

[2] MSU-DOE Plant Research Laboratory, Michigan State University, East Lansing, MI, USA

[3] Environmental Genomics and Systems Biology Division, Lawrence Berkeley National Laboratory, Berkeley, CA, USA

[4] Molecular Biophysics and Integrated Bioimaging Division, Lawrence Berkeley National Laboratory, Berkeley, CA, USA

[5] Department of Biochemistry and Molecular Biology, Michigan State University, East Lansing, MI, USA

[6] Pritzker School of Molecular Engineering, University of Chicago, Chicago, IL, USA

[7] Institute for Biophysical Dynamics, University of Chicago, Chicago, IL, USA

[8] Chan-Zuckerberg Biohub Chicago, Chicago, IL, USA

* e-mail correspondence: asquires@uchicago.edu

# Contents

# Supplementary Notes

## Supplementary Note S1: Sequence and structural analysis

We compared the sequences of ApcA and ApcB subunits from tri-PBS and penta-PBS (**Fig. S1**) as well as sequences of tri-OCP and penta-OCP (**Fig. S2**). Sequences were aligned and visualized using clustalX (1). The common binding motif in the known tri-OCP binding sites on the tri-PBS consists of two ApcA and one ApcB subunits. We identified all the sites where this motif is present in both the tri-PBS and penta-PBS. We aligned all possible binding sites to the high-resolution complex of tri-OCP with the binding pocket from the top cylinder of tri-PBS (PDB: 8TPJ). After aligning, we computed the center of mass and radius of gyration of the NTD of the OCP which we used to place spheres at all the binding sites for easier visualization. At each binding site, we also computed the orientation of OCP linker near the NTD using Cα atoms from residues 171 and 178 of the tri-OCP which is shown as a grey arrow (see **Fig. S5** for visualization of the alignment procedure). Structural alignment was performed with the PyMOL (2) super command, using only Cα residues.

## Supplementary Note S2: Data Analysis

**Filtering the single-molecule data:** In the scatter heatmaps shown in **Figs. 2-3** of the main text, the data undergoes a few cleanup steps to better visualize the features of interest. The cleanup procedure is as follows.

As described in the methods, the time-tagged photon-by-photon data are binned into 10-ms bins for each detection channel. The binned brightness trace from each channel is passed to the change-point finding algorithm which detects the bins where a change in intensity occurs. These changepoints are combined for both detection channels. The bins between two changepoints are considered a 'level' and correspond to an interval of time when the intensity in both channels is constant. Photons from each level are separated into groups of 500 photons. For each 500-photon group, we calculate the total brightness, lifetime using photons from the red (or green) detection channel, and red ratio.

The scatter heatmaps created from all 500-photon groups (hereafter called bins) for each experiment are shown in **Figs. S9-S10**. To create the plots for the figures in main text, we remove bins from molecules that are likely oligomers, bins during photobleaching, and bins from molecules that were trapped for less than 0.5 seconds.

When trapping phycobilisomes bound to OCP, higher excitation intensities are required to measure single molecules. At these intensities, phycobilisomes that are not bound to OCP can quickly begin to photobleach. To remove the bins during photobleaching, we use the mean and standard deviation of brightness and red lifetime of unquenched phycobilisomes to create an ellipse. In the lifetime vs. brightness projection, bins that are in this ellipse are likely to be unquenched phycobilisomes. For each trace, if any of the bins are inside this ellipse, we only keep the bins that are inside the ellipse and eliminate any bins are outside the ellipse for that trace (see example event in **Fig. S15**).

To remove oligomers, we use the red lifetime vs. total brightness data from trapping individual phycobilisomes (**Figs. S9a and S10a**) to determine a slanted line (dashed yellow) that separates monomers from oligomers. We exclude all bins from traces that includes any bins below the slanted line (see example event **Fig. S16**).

**Data analysis for titration experiments:** Data from red lifetime vs. red ratio projections of separate experiments (**Fig. S18**) performed with different phycobilisome to OCP ratios were analyzed to determine the number of molecules in Q1 or Q2 as fraction of total quenched molecules. 95% confidence ellipses centered on Q1, Q2, or unquenched populations in the red lifetime vs. red ratio projections were used to assign each trapped molecule into the appropriate photophysical states. Any trapped molecule with at least one 500-photon bin in the U state was classified as unquenched. The rest of the trapped molecules were assigned to either Q1 or Q2 if they visited a quenched state; molecules with bins in both Q1 and Q2 photophysical states were assigned to the state that was visited first.

## Supplementary Note S3: Photon-by-photon Monte Carlo simulations

A compartmental model (3, 4) was used for photon-by-photon Monte Carlo simulations. The compartmental model places groups of rod or core pigments into distinct compartments and uses data from time-resolved fluorescence spectroscopy to describe the flow of excitation energy within the compartments. The existing tri-PBS compartmental model architecture was modified to reflect the architecture and connectivity of the penta-PBS by adding two additional flanking core half-cylinders with two additional rods. We used the rates given in ref. (5) for the rod-to-core exciton transfer in the penta-PBS and the rates within the core compartments were taken directly from the tri-PBS model (**Figs. S22-S23**).

During the simulation, a single compartment is populated with an excitation based on its absorption probability. The absorption probabilities for each compartment are shown in **Figs. S22-S23**. After exciting a single compartment, the excitation randomly travels to nearby compartments based on the excitation transfer rates with a time step of 100 fs. During each time step, the excitation can stay at the same compartment, transfer to a new compartment, or emit in the form of fluorescence with a radiative decay rate of 0.6 $ns^{-1}$ and quantum yield of 50%. To simulate the quenching of phycobilisomes, we placed an OCP compartment with a variable quenching rate and connected to different core compartments. We simulated $10^5$ absorption events and recorded the total number of emitted photons, time upon emission, as well as the delay time between the absorption and emission events for further processing.

The emitted photons from the simulation were grouped into 200 photon bins. For each bin, we calculated the brightness and lifetime which are then averaged together. We normalized the simulated brightnesses using the number of emitted photons in our simulations and the experimentally observed brightness for unquenched phycobilisomes. The simulated lifetime is based on a single-exponential fit of the data.

## Supplementary Note S4: Additional discussion of penta-PBS binding sites for penta-OCP

Another possible binding site on the top cylinder involves T3 which could form a pair with the E5' site (**Fig. S26a**). However, this would also require a different linker orientation. Additionally, the T3 site is not occupied in the tri-PBS. In the penta-PBS, the T3 is in close proximity to the E' cylinder as well as the R3' rods which makes it a less likely candidate. The other possible pairs involve the B9 site which could form a pair with the E2 or E5 sites (**Figs. S26b-c**). However, the B9 site has the largest RMSD from the known OCP-binding sites on the tri-PBS. Also, its pairing with either E2 or E5 would place the bound NTDs of OCP$^R$ very close to each other, especially when paired with E2 (49 Å) compared to the distance between the NTD centers of mass in tri-PBS (85 Å).

# Supplementary Figures

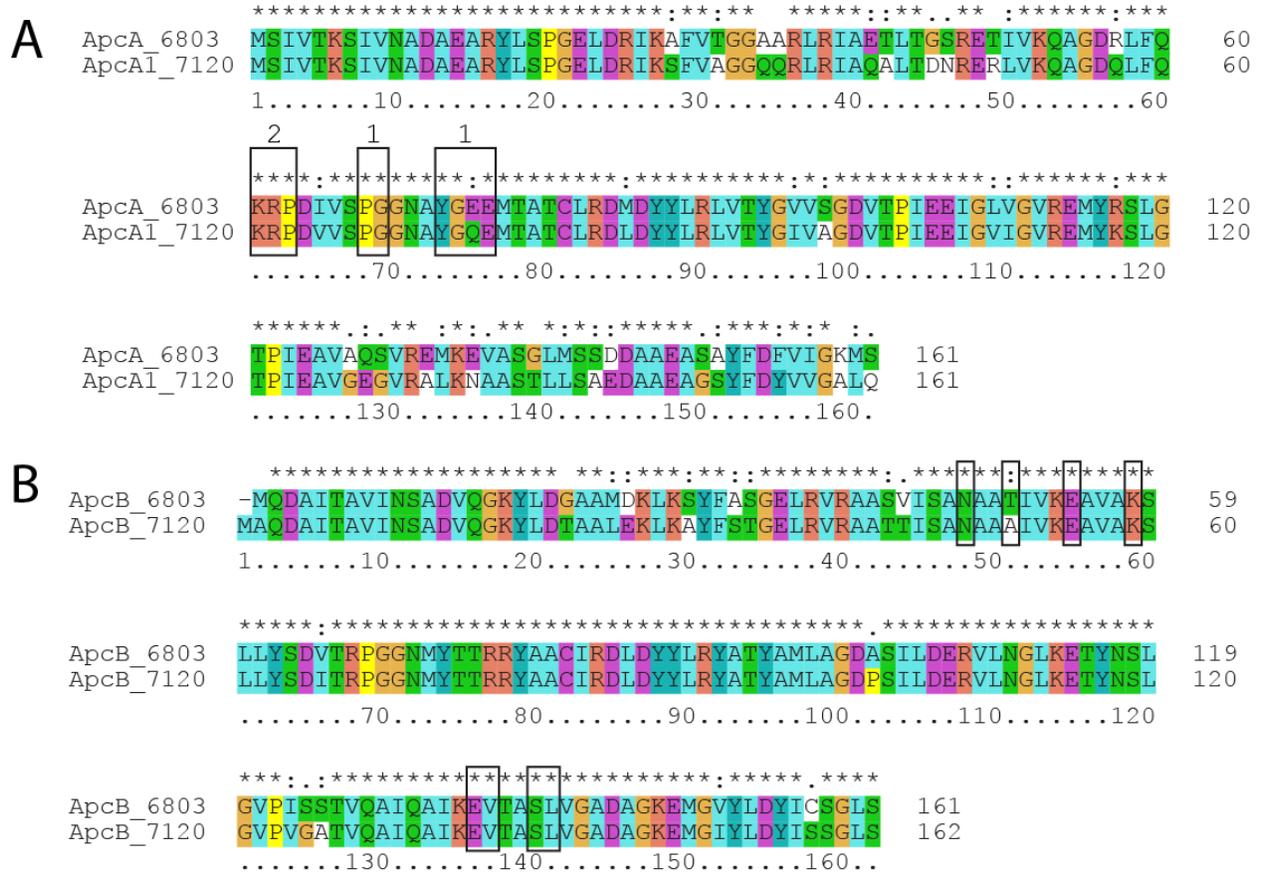

**Fig. S1. Sequence alignment of ApcA and ApcB from tri-PBS and penta-PBS**

Sequence alignment of (a) ApcA (Slr2067 and Alr0021) and (b) ApcB (Slr1986 and Alr0022) from tri-PBS and penta-PBS was performed. OCP$^R$(NTD)-PBS complex residues (within 3 Å of each other) are highlighted with boxes (numbers above boxes in (a) refer to the two different ApcA subunits). Conserved residues are colored according to their properties; an asterisk above sequence indicates identity, a colon homology and a period similarity.

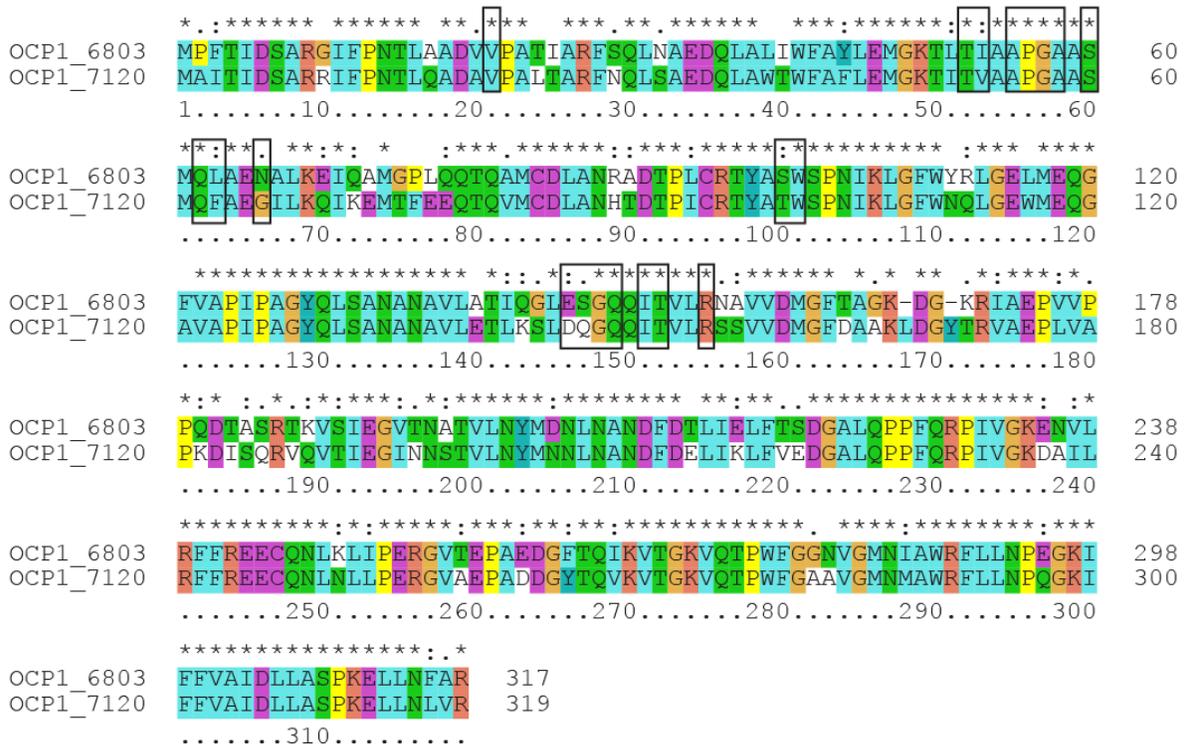

**Fig. S2. Sequence alignment of tri-OCP and penta-OCP**

Sequence alignment of the tri-OCP (Slr1963) and penta-OCP (All4941) was performed. OCP$^R$(NTD)-PBS complex residues (within 3 Å of each other) are highlighted with boxes. Conserved residues are colored according to their properties; an asterisk above sequence indicates identity, a colon homology and a period similarity. The N-terminal domain, the linker region, and the C-terminal domain are 84%, 74%, and 92% conserved, respectively.

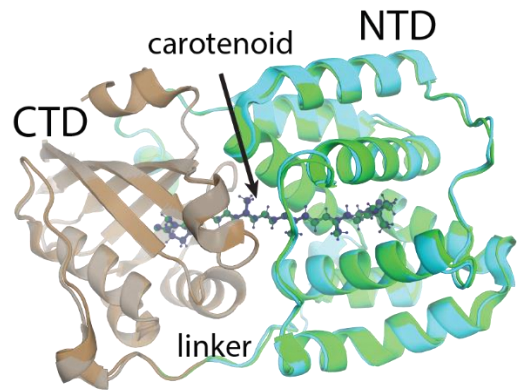

**Fig. S3. Overlay of tri-OCP$^O$ and penta-OCP$^O$ structures**

Structural superposition of *Synechocystis* PCC. 6803 OCP$^O$ (tri-OCP$^O$) (NTD: lime, CTD: grey; PDB: 3MG1 (6)) and *Anabaena* PCC. 7120 OCP$^O$ (penta-OCP$^O$) (NTD: cyan, CTD: tan; PDB: 5HGR (7)) in the inactive state with an RMSD of 0.341 Å.

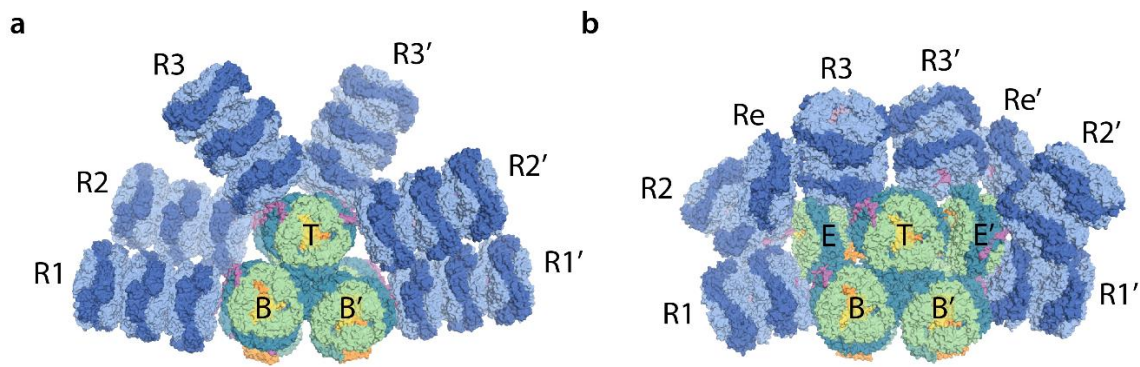

**Fig. S4. Structures of tri-PBS and penta-PBS with labeled rods and core cylinders**

Cryo-EM structure of (a) tri-PBS (Domínguez-Martín et al. (8)) and (b) penta-PBS (PDB: 7EYD; Zheng et al. (9)) with the rods and core cylinders labeled with identifiers used in the SI table 1. Labels denoted by a prime (´) denote structurally equivalent rod and core cylinders after a 180° horizontal rotation.

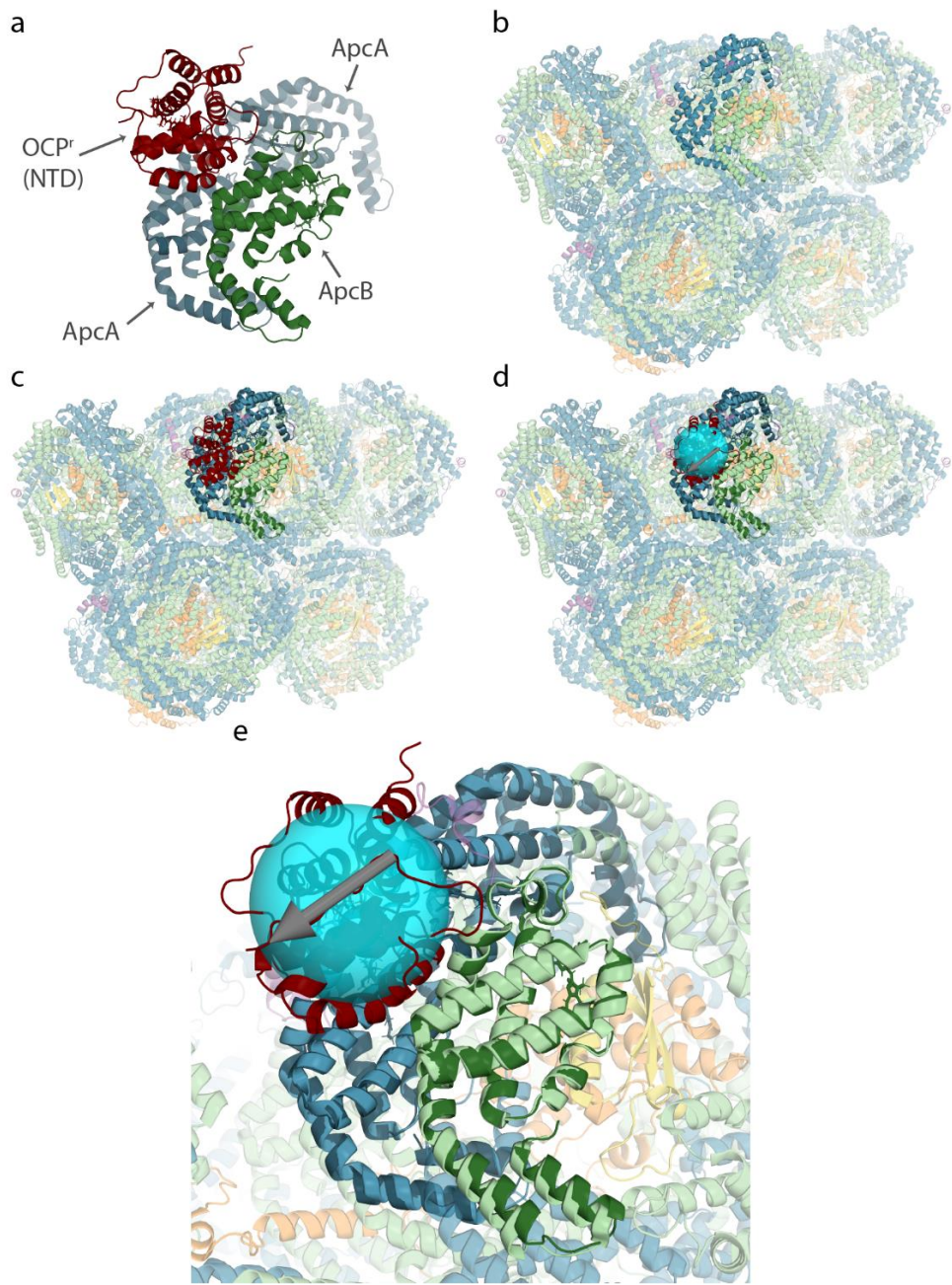

**Fig. S5. Procedure for identifying possible OCP$^R$(NTD) binding sites on penta-PBS**

We used the high-resolution structure of OCP bound to the top cylinder of tri-PBS (PDB: 8TPJ) to align OCP$^R$(NTD) to the possible binding sites on the penta-PBS. a) The binding pocket consists of two ApcA and one ApcB subunit. b) We identify all sites where this binding pocket appears on the penta-PBS. One instance of this binding motif on the top cylinder of penta-PBS is shown. c) The known tri-PBS binding motif is aligned to a possible penta-PBS binding pocket. d) The location of OCP$^R$(NTD) after alignment is used to calculate a center-of-mass, radius of gyration, and linker orientation which are shown as a blue sphere and grey arrow for easier visualization. e) A close up look at the alignment results.

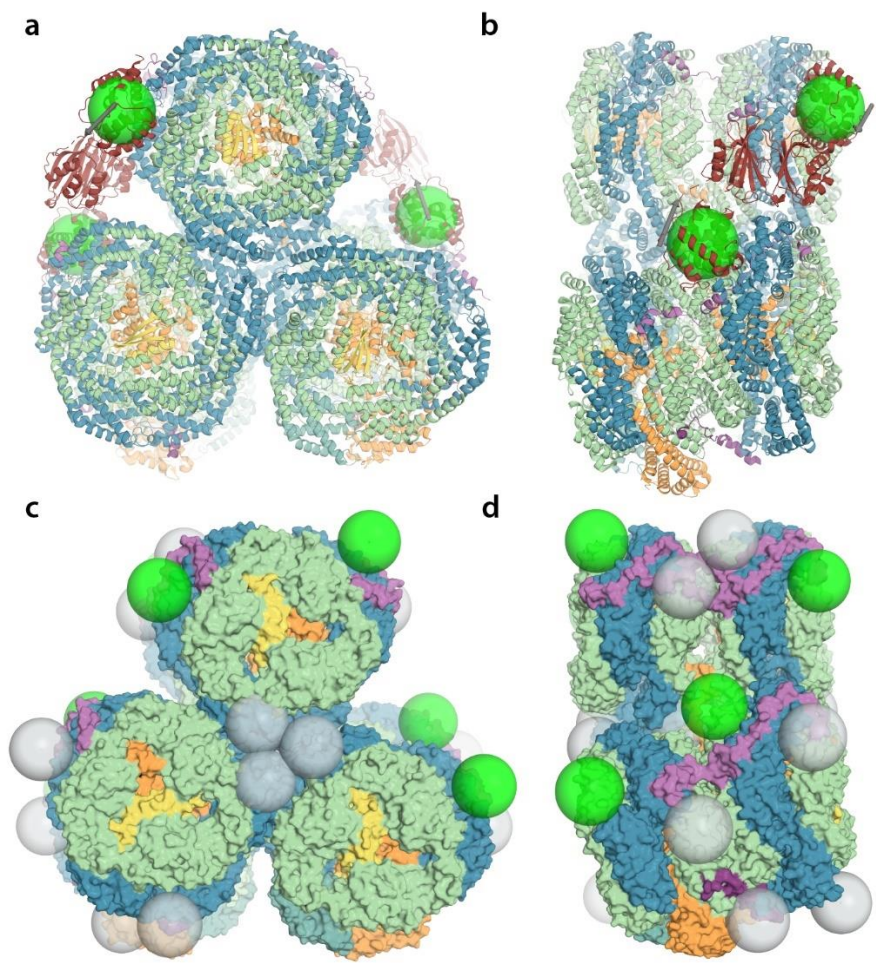

**Fig. S6. The known and accessible tri-OCP$^R$(NTD) binding sites on tri-PBS**

The front (a) and sideview (b) of tri-OCP dimers (red) bound to the tri-PBS core are shown (PDB: 7SC9). The lime spheres are centered at the centers of mass of the NTD and have a radius equal to the radius of gyration of the NTD. The grey arrows indicate the orientation of the linker connecting the NTD and CTD of each bound OCP. We identified all sites on the tri-PBS core where the binding motif appears and aligned them with the binding motif from the known binding site at the top cylinder (c, d). Location of the NTD of OCP$^R$ after alignment is shown as a sphere with green spheres representing accessible binding sites and grey spheres representing binding site that are structurally blocked.

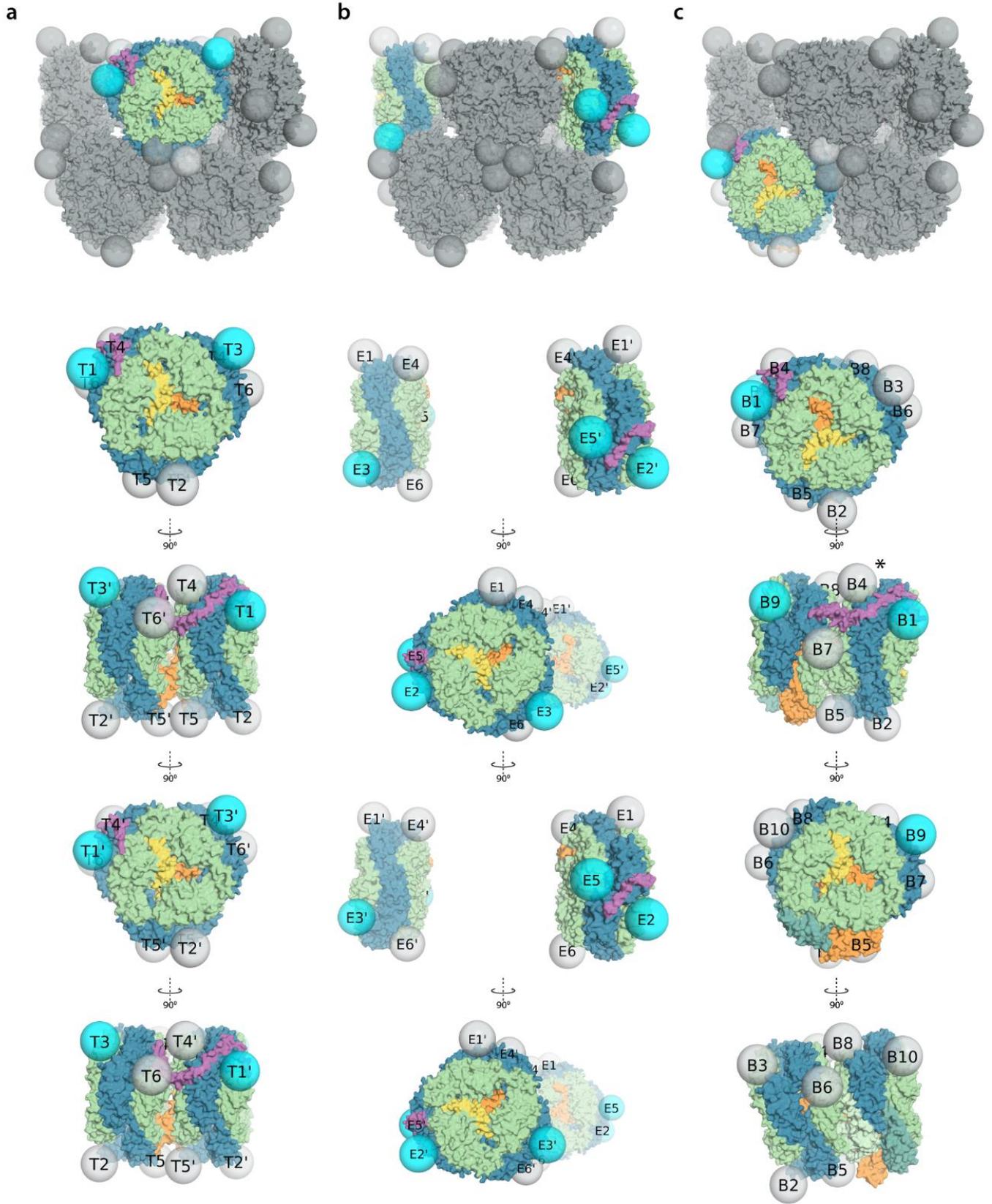

**Fig. S7. (Previous page) Naming conventions and possible binding sites for OCP$^R$(NTD)**

Each possible OCP$^R$(NTD) binding site is labeled with a unique identifier which is used in tables S1-S4. Labels denoted by a prime (´) denote structurally equivalent sites after a 180° horizontal rotation. a) The top (T) cylinder has 12 possible OCP binding sites. The 6 binding sites on the front hexamer (T1 to T6) have structurally similar analogs on the back hexamer (T1' to T6'). b) The two extra flanking core cylinders in penta-PBS have 6 possible binding sites each. The 6 binding sites on the E cylinder (E1 to E6) are structurally equivalent to the 6 binding sites on the E' cylinder (E1' to E6'). c) The bottom cylinder (B) has 10 possible binding sites. The binding site B4 (marked with an asterisk) is a known binding site in tri-PBS but is blocked by the E cylinder in the penta-PBS. The second bottom cylinder (B') also has 10 possible binding sites (not shown) which are analogous to the binding sites on the B cylinder. Note that tri-PBS sites use the same nomenclature, except that tri-PBS lacks the E cylinders.

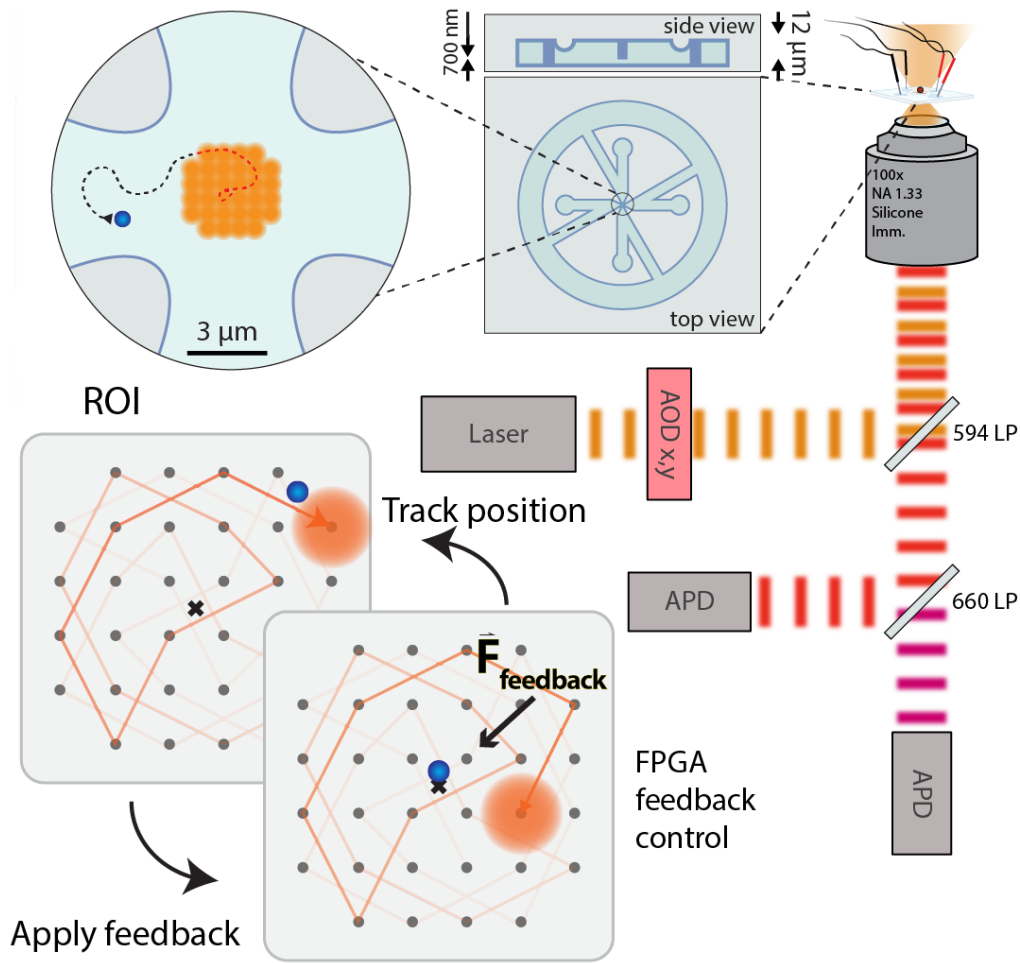

**Fig. S8. ABEL trap setup**

The Anti-Brownian ELectrokinetic (ABEL) trap scans a confocal beam (594 nm) in a knight's tour pattern. Upon detecting a photon from a freely diffusing fluorescent molecule in the microfluidic cell, a voltage proportional to the distance from the center of the region of interest (ROI) is applied to keep the molecule in the center of the trap. This process is repeated until the molecule photobleaches, diffuses out of the trap, or gets kicked out of the ROI after being trapped for a set amount of time. In our setup, the fluorescence from the trapped molecule is split using a 660 long pass (LP) dichroic into "red" (emission above 600 nm) and "green" (emission below 660 nm) channels to gain spectral information.

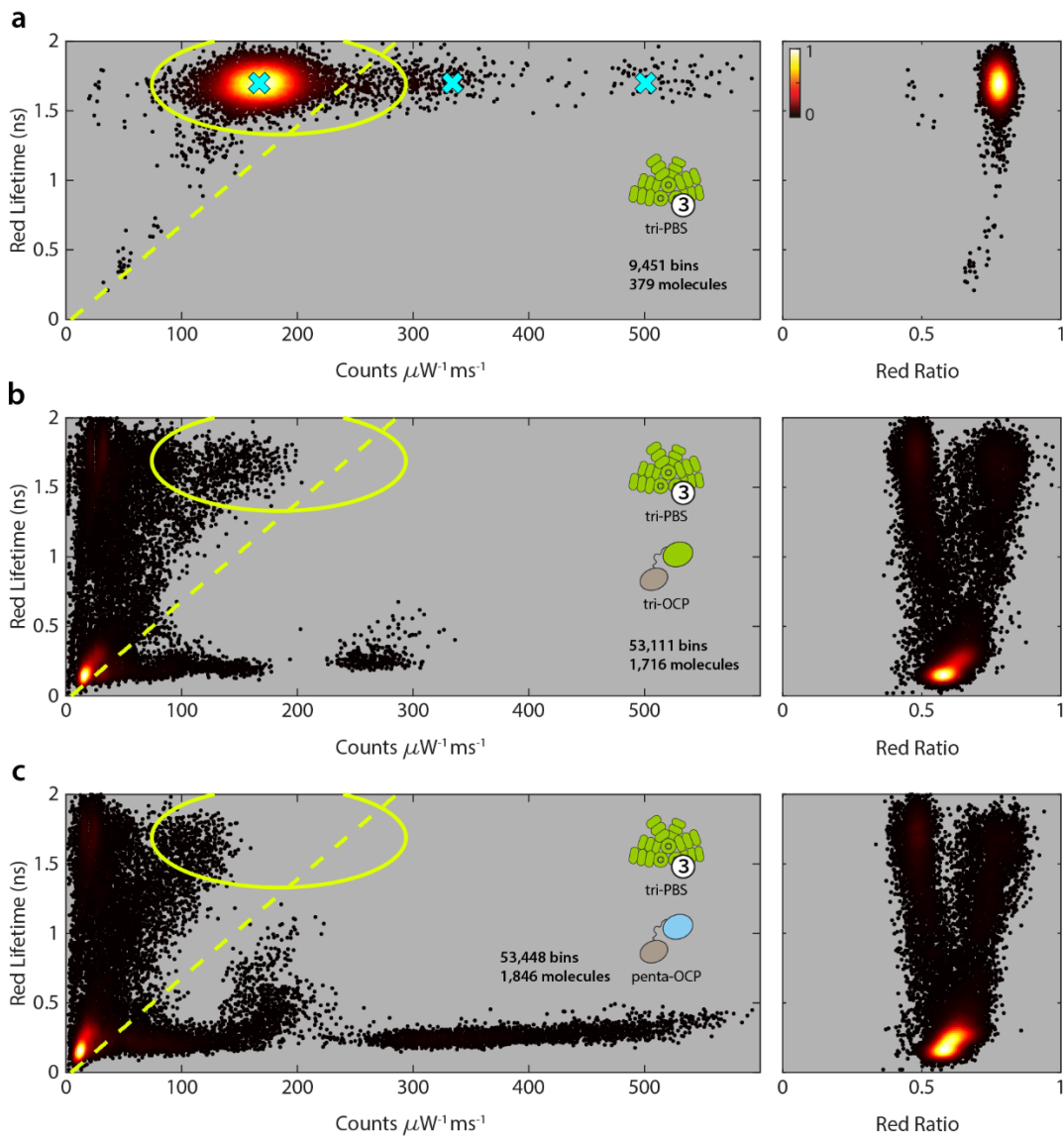

**Fig. S9. Unfiltered single-molecule data for quenched and unquenched tri-PBS.**

Scatter heatmaps showing red channel lifetime vs. normalized brightness (left) or red ratio (right) projections for (a) tri-PBS (b) tri-PBS with tri-OCP and (c) tri-PBS with penta-OCP without any filtering of trapping events. Each dot represents a spectroscopic parameter calculated for a group of 500 photons (bin). The cyan crosses in (a) are integer multiples of the brightness of the unquenched phycobilisome monomers. The dotted line is used to eliminate oligomers by removing any trace containing bins that fall below the line. The ellipse mostly contains bins from unquenched phycobilisomes and is used to eliminate bins during photobleaching by only keeping the bins inside the ellipse for a trace that contains bins both inside and outside the ellipse.

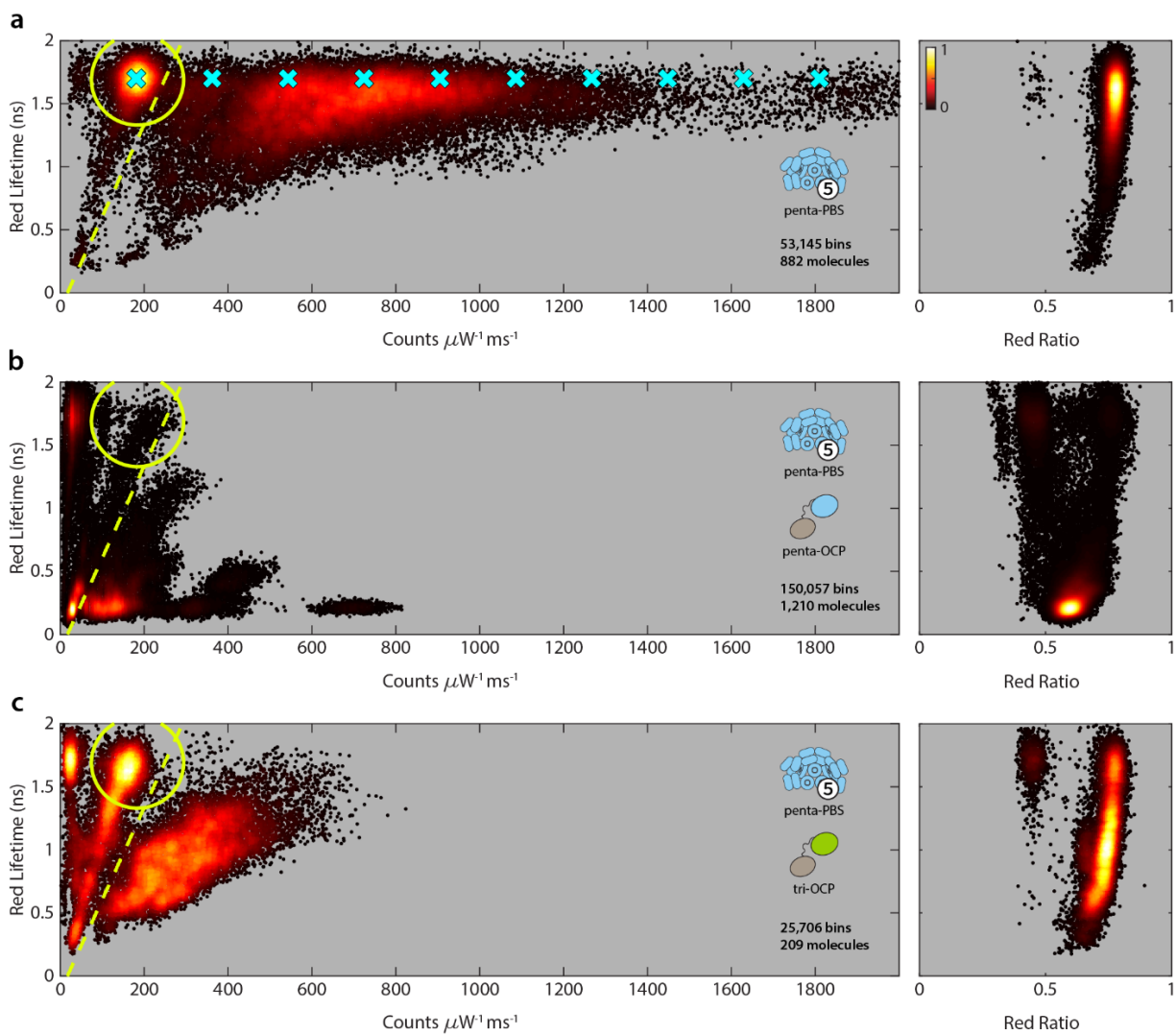

**Fig. S10. Unfiltered single-molecule data for quenched and unquenched penta-PBS**

Scatter heatmaps showing red channel lifetime vs. normalized brightness (left) or red ratio (right) projections for (a) penta-PBS (b) penta-PBS with penta-OCP and (c) penta-PBS with tri-OCP without any filtering of trapping events. Each dot represents a spectroscopic parameter calculated for a group of 500 photons (bin). The cyan crosses in (a) are integer multiples of the brightness of the unquenched phycobilisome monomers. The dotted line is used to eliminate oligomers by removing any trace containing bins that fall below the line. The ellipse mostly contains bins from unquenched phycobilisomes and is used to eliminate bins during photobleaching by only keeping the bins inside the ellipse for a trace that contains bins both inside and outside the ellipse.

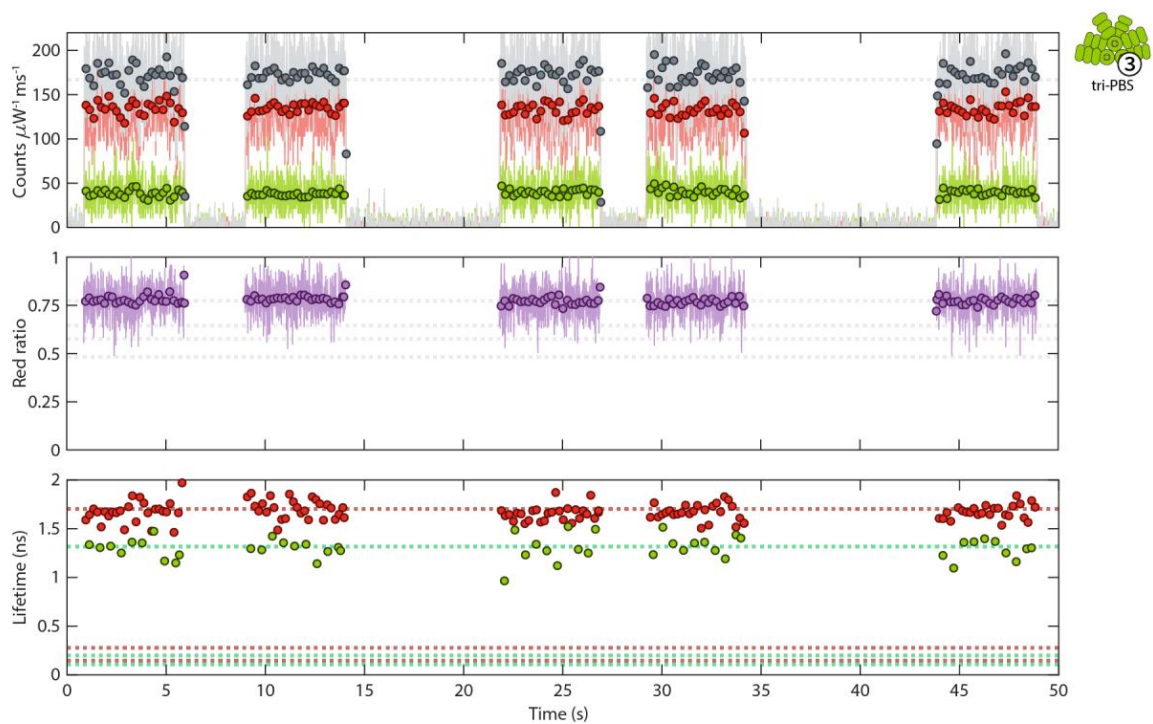

**Fig. S11. Example trapping traces of tri-PBS at low excitation intensity**

Single tri-PBS exhibit consistent spectroscopic parameters including brightness (top), red ratio (middle), and red/green lifetimes (bottom) at low excitation intensity (0.1 W/cm$^2$). Parameters determined for 10 ms bins (solid lines) or 500 photon groups (markers) are plotted. The top panel shows the total brightness (grey), red channel brightness (red), and green channel brightness (green). The dotted line at 167 counts µW$^{-1}$ ms$^{-1}$ is the average total brightness of tri-PBS. The dotted lines in middle and bottom panel refer to key photophysical states observed for quenched or unquenched tri-PBS at low and/or high excitation intensity. In the middle panel, the dotted lines refer to red ratio of unquenched tri-PBS (0.775), Q1 (0.645), Q2 (0.577), and rods (0.482). In the bottom panel, each marker corresponds to fitted lifetimes determined from groups of 500 red channel photons (red markers) or 500 green channel photons (green markers). The red dotted lines refer to the average red lifetime of unquenched tri-PBS (1.70), Q1 (0.28), and Q2 (0.15). The green dotted lines refer to average green lifetime of unquenched tri-PBS (1.32), Q1 (0.20), and Q2 (0.11). At the low excitation intensity, only the pristine, unquenched tri-PBS is observed. Note that in this experiment, the trap kicks out the particle after 5 seconds of trapping in order to increase the number of distinct particles measured during the experiment.

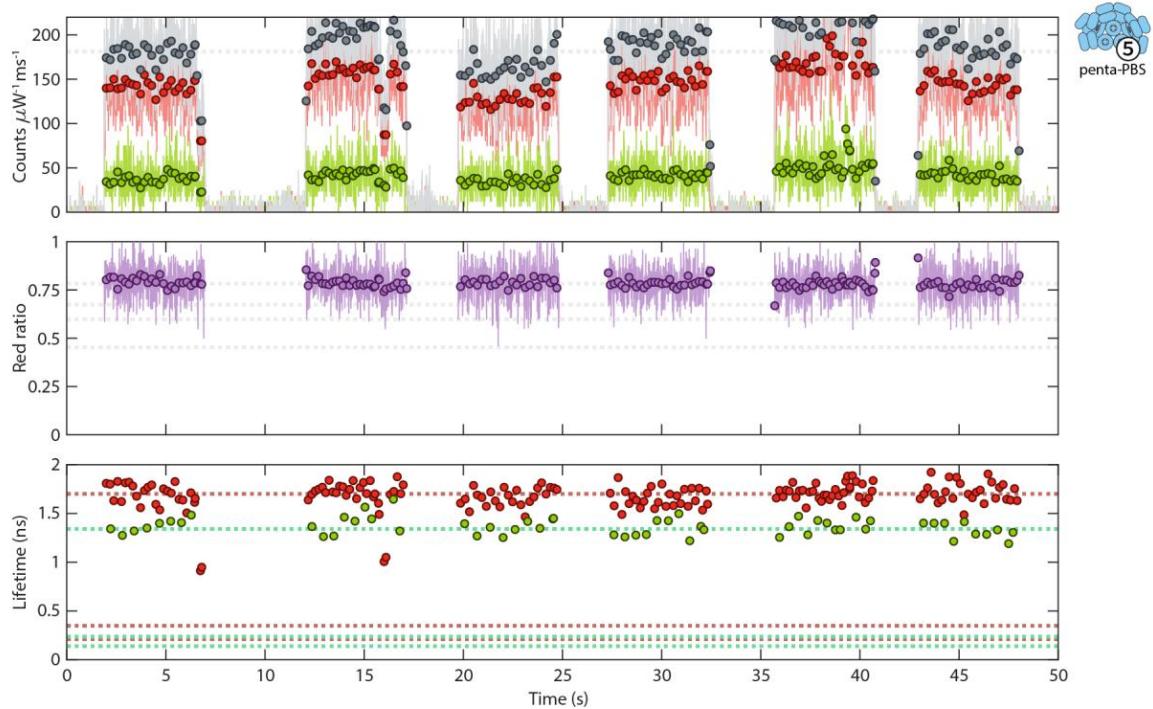

**Fig. S12. Example trapping traces of penta-PBS at low excitation intensity**

Single penta-PBS exhibit consistent spectroscopic parameters including brightness (top), red ratio (middle), and red/green lifetimes (bottom) at low excitation intensity (0.1 W/cm$^2$). Parameters determined for 10 ms bins (solid lines) or 500 photon groups (markers) are plotted. The top panel shows the total brightness (grey), red channel brightness (red), and green channel brightness (green). The dotted line at 181 counts µW$^{-1}$ ms$^{-1}$ is the average total brightness of penta-PBS. The dotted lines in middle and bottom panel refer to key photophysical states observed for quenched or unquenched penta-PBS at low and/or high excitation intensity. In the middle panel, the dotted lines refer to red ratio of unquenched penta-PBS (0.783), Q1 (0.673), Q2 (0.597), and rods (0.453). In the bottom panel, each marker corresponds to fitted lifetimes determined from groups of 500 red channel photons (red markers) or 500 green channel photons (green markers). The red dotted lines refer to the average red lifetime of unquenched penta-PBS (1.70), Q1 (0.35), and Q2 (0.21). The green dotted lines refer to average green lifetime of unquenched penta-PBS (1.34), Q1 (0.24), and Q2 (0.14). At the low excitation intensity, only the pristine, unquenched penta-PBS is observed. Note that in this experiment, the trap kicks out the particle after 5 seconds of trapping in order to increase the number of distinct particles measured during the experiment.

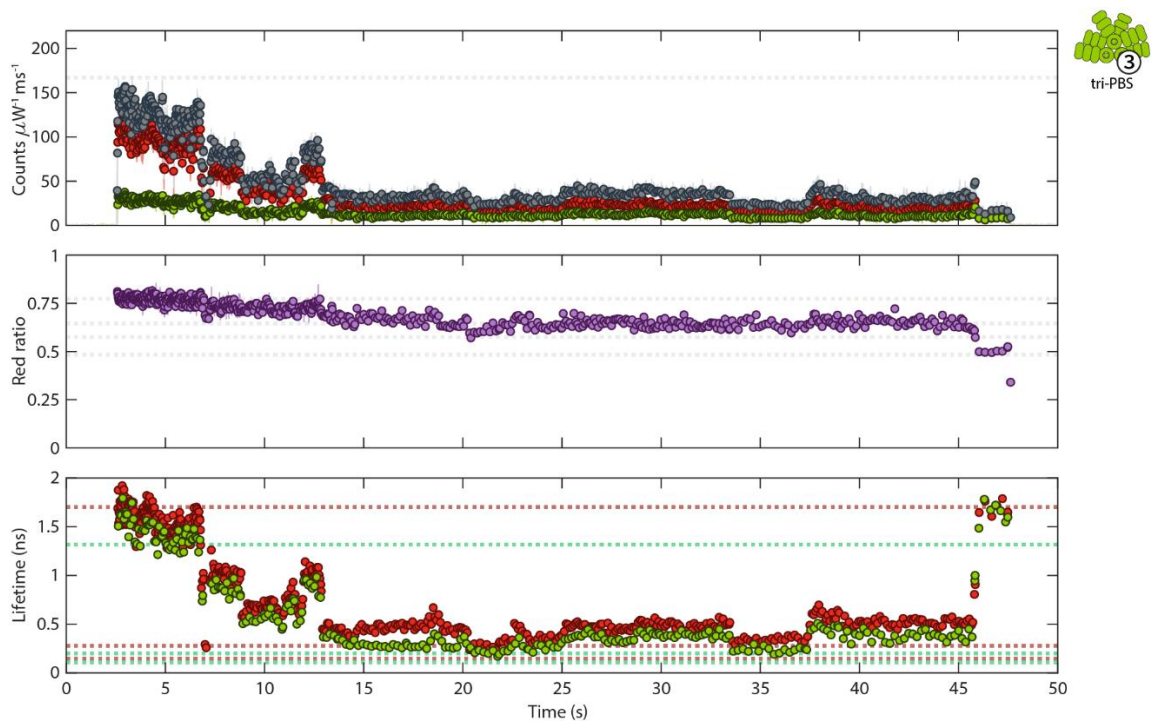

**Fig. S13. Example trapping traces of tri-PBS at high excitation intensity**

Single tri-PBS without any OCP present quickly photobleaches at high excitation intensity (1.2 W/cm$^2$). Parameters determined for 10 ms bins (solid lines) or 500 photon groups (markers) are plotted. The top panel shows the total brightness (grey), red channel brightness (red), and green channel brightness (green). The dotted line at 167 counts µW$^{-1}$ ms$^{-1}$ is the average total brightness of unquenched tri-PBS without any photobleaching. The dotted lines in middle and bottom panel refer to key photophysical states observed for quenched or unquenched tri-PBS at low and/or high excitation intensity. In the middle panel, the dotted lines refer to red ratio of unquenched tri-PBS (0.775), Q1 (0.645), Q2 (0.577), and rods (0.482). In the bottom panel, each marker corresponds to fitted lifetimes determined from groups of 500 red channel photons (red markers) or 500 green channel photons (green markers). The red dotted lines refer to the average red lifetime of unquenched tri-PBS (1.70), Q1 (0.28), and Q2 (0.15). The green dotted lines refer to average green lifetime of unquenched tri-PBS (1.32), Q1 (0.20), and Q2 (0.11). Similar excitation intensities were used to measure the quenched phycobilisomes. However, unlike the quenched phycobilisomes, unquenched phycobilisomes photobleach at these intensities as shown by the decrease in brightness (top), red ratio (middle), and lifetime (bottom) upon trapping.

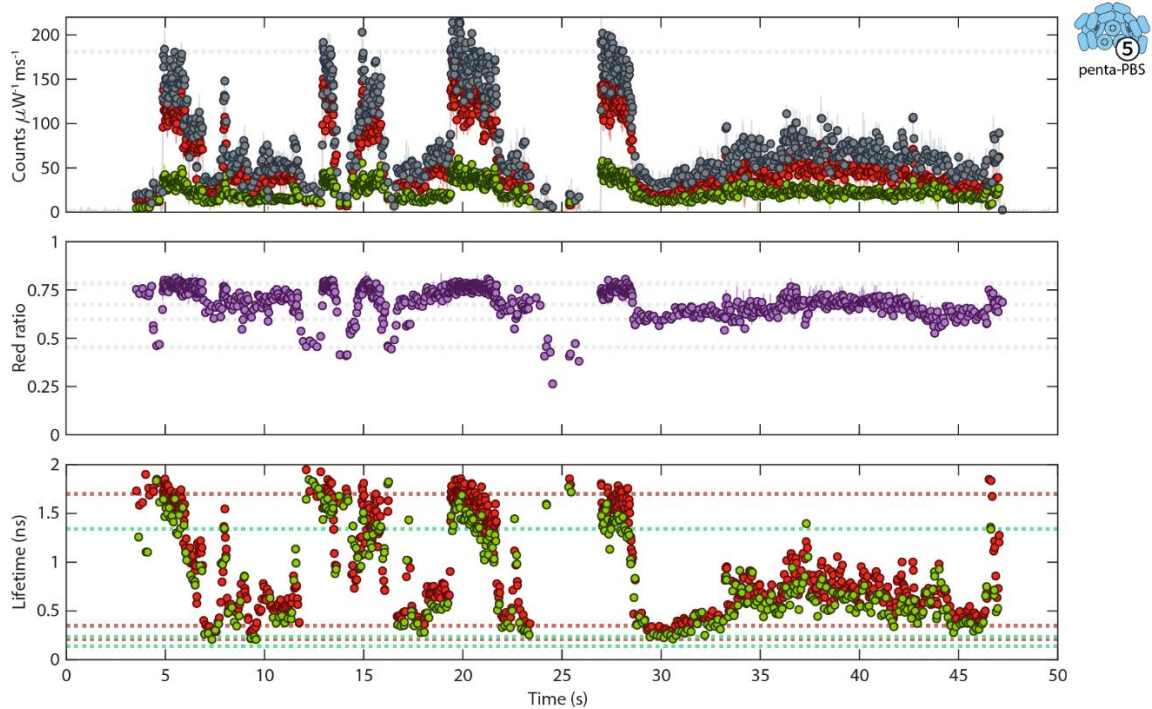

**Fig. S14. Example trapping traces of penta-PBS at high excitation intensity**

Single penta-PBS without any OCP present quickly photobleaches at high excitation intensity (1.1 W/cm$^2$). Parameters determined for 10 ms bins (solid lines) or 500 photon groups (markers) are plotted. The top panel shows the total brightness (grey), red channel brightness (red), and green channel brightness (green). The dotted line at 181 counts µW$^{-1}$ ms$^{-1}$ is the average total brightness of penta-PBS without any photobleaching. The dotted lines in middle and bottom panel refer to key photophysical states observed for quenched or unquenched penta-PBS at low and/or high excitation intensity. In the middle panel, the dotted lines refer to red ratio of unquenched penta-PBS (0.783), Q1 (0.673), Q2 (0.597), and rods (0.453). In the bottom panel, each marker corresponds to fitted lifetimes determined from groups of 500 red channel photons (red markers) or 500 green channel photons (green markers). The red dotted lines refer to the average red lifetime of unquenched penta-PBS (1.70), Q1 (0.35), and Q2 (0.21). The green dotted lines refer to average green lifetime of unquenched penta-PBS (1.34), Q1 (0.24), and Q2 (0.14). Similar excitation intensities were used to measure the quenched phycobilisomes. However, unlike the quenched phycobilisomes, unquenched phycobilisomes photobleach at these intensities as shown by the decrease in brightness (top), red ratio (middle), and lifetime (bottom) upon trapping. The sudden increases in brightness during the first trapping event are likely due to replacement events.

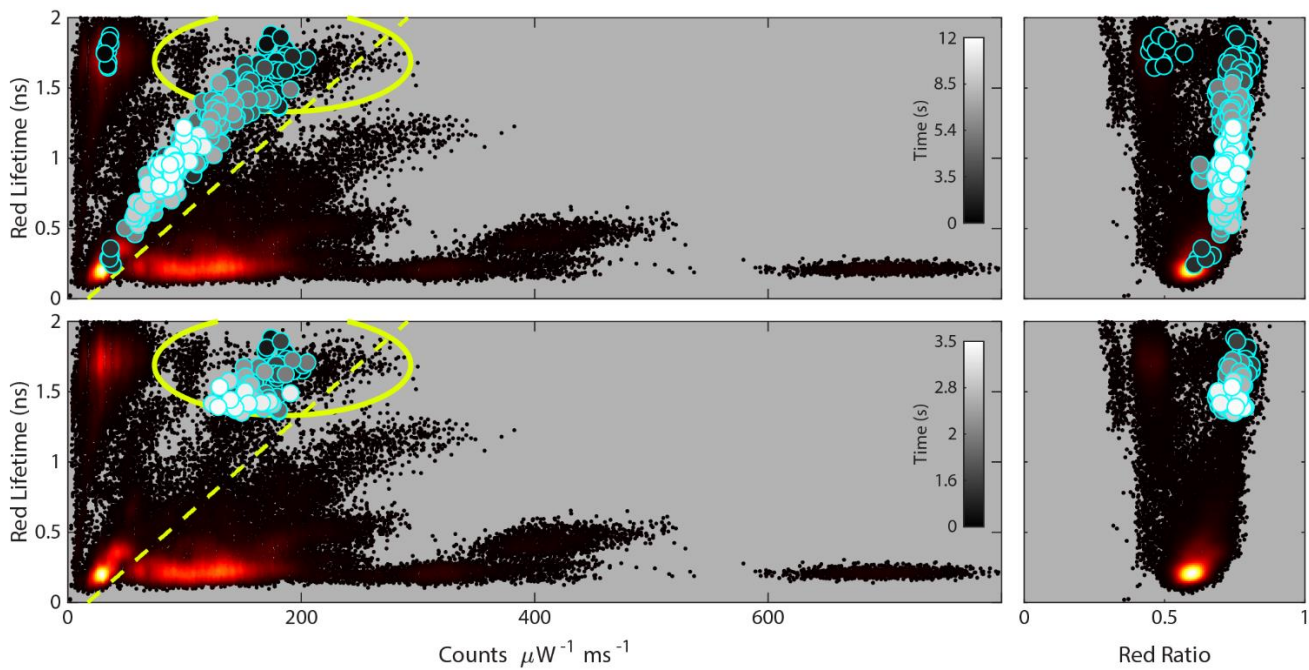

**Fig. S15. Heatmaps showing penta-PBS quenched with penta-OCP with a photobleaching example event before and after filtering**

Red lifetime vs. brightness (left) and red lifetime vs. red ratio (right) scatter heatmaps (yellow-red-black) show the complete dataset obtained for penta-PBS quenched with penta-OCP. Markers with cyan outline show an example event where a single penta-PBS enters the trap and begins photobleaching. The example trace is colored according to measurement time with beginning of the trace shown in black and end of trace shown in white (as indicated by the color bar). The full trace is shown in the top panel. To filter out the data during photobleaching that would overlap with the true OCP-quenched states in the lower left of the graph, we only keep the bins that are in the ellipse representing unquenched PBS and eliminate all other bins as shown in the bottom panel.

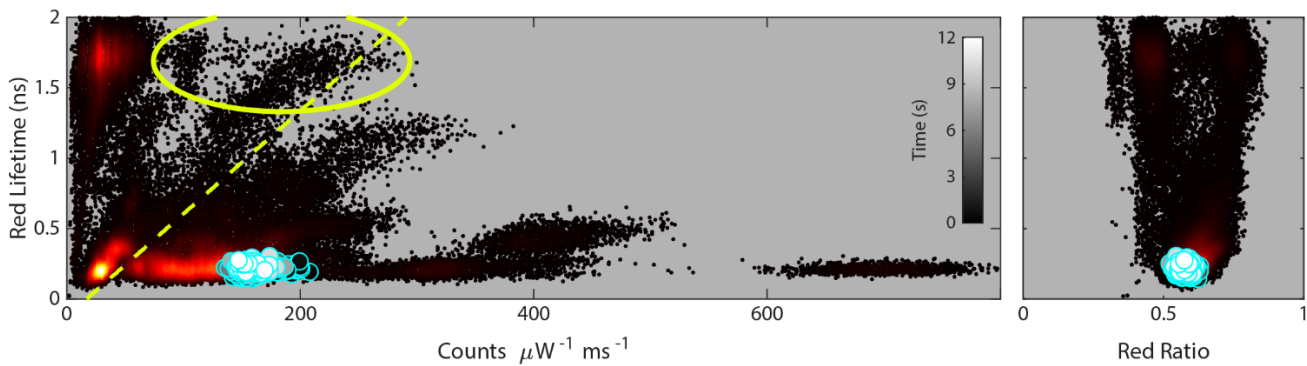

**Fig S16. Heatmaps showing penta-PBS quenched with penta-OCP with an eliminated example event**

Red lifetime vs. brightness (left) and red lifetime vs. red ratio (right) scatter heatmaps (yellow-red-black) show the complete dataset obtained for penta-PBS quenched with penta-OCP. Markers with cyan outline show an example event that is likely a quenched phycobilisome oligomer. The example trace is colored according to measurement time with beginning of the trace shown in black and end of trace shown in white (as indicated by the color bar). Since it has a low lifetime, it is likely quenched by the penta-OCP. However, based on its brightness it likely not a monomer (bins below the dotted line) and therefore all bins are eliminated from the filtered data.

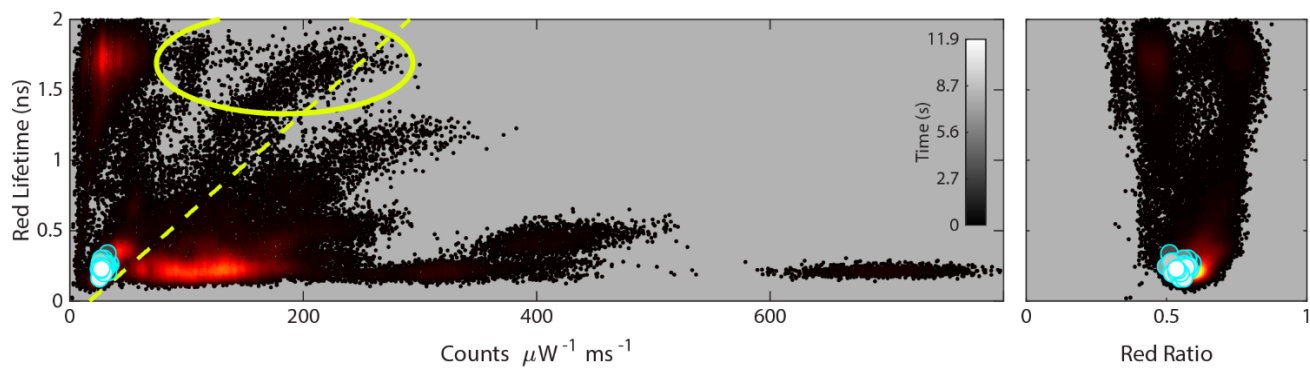

**Fig. S17. Heatmaps showing penta-PBS quenched with penta-OCP with an unfiltered example event**

Red lifetime vs. brightness (left) and red lifetime vs. red ratio (right) scatter heatmaps (yellow-red-black) show the complete dataset obtained for penta-PBS quenched with penta-OCP. Markers with cyan outline show an example event that is quenched (no bins in the ellipse) and likely not an oligomer (bins do not go below the dotted line). The example trace is colored according to measurement time with beginning of the trace shown in black and end of trace shown in white (as indicated by the color bar). All bins from this trapping event are kept during data filtration.

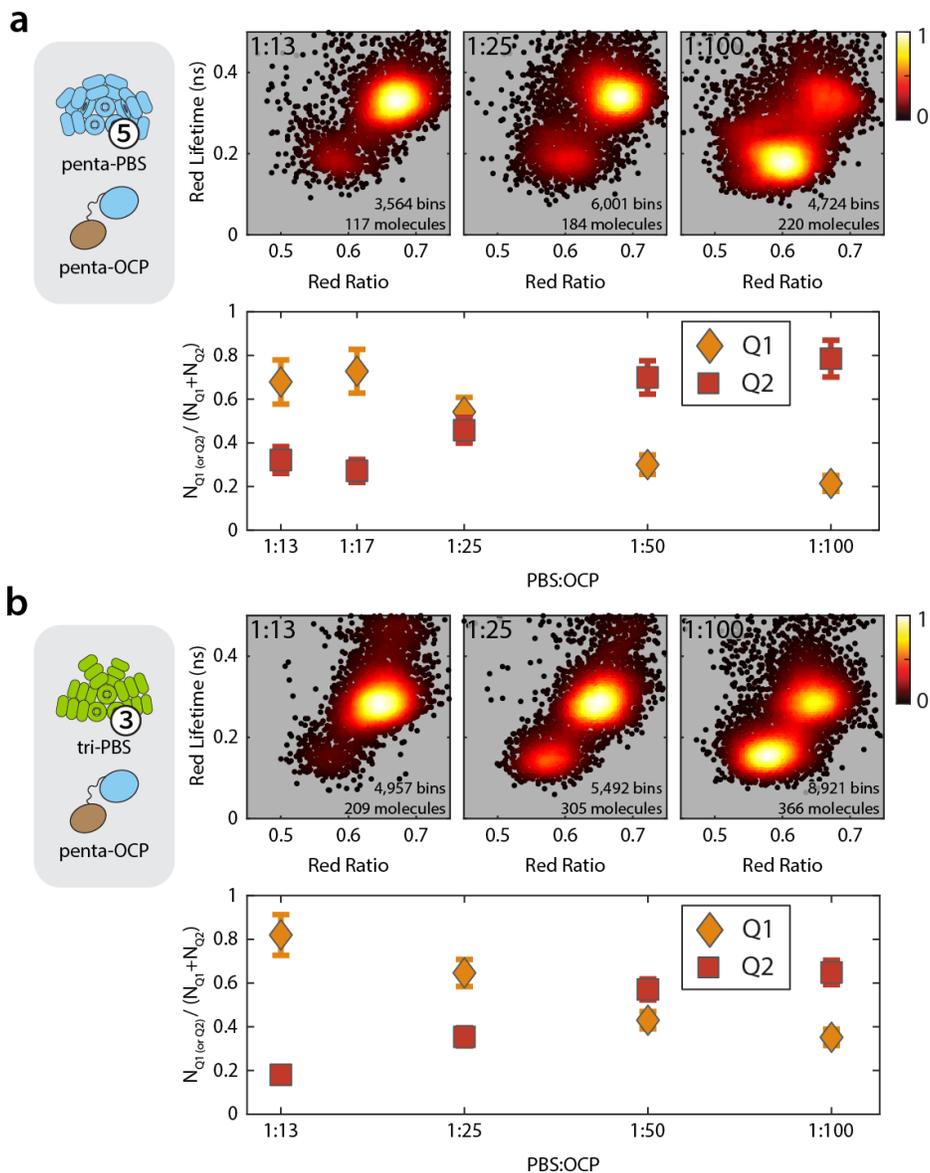

**Fig. S18. Shift in the proportion of molecules in Q1 vs. Q2 state upon OCP titration**

a) Penta-PBS titrated by the penta-OCP. The top three panels show red lifetime vs. red ratio scatter heatmaps from three separate experiments with 1:13, 1:25, and 1:100 phycobilisome to OCP ratios. The bottom panel in (a) quantifies the scatter heatmaps by finding out the number of molecules in Q1 and Q2 states. The red and orange markers show the fraction of Q2 and Q1 molecules measured during a 2-hour trapping experiment, respectively. b) Similarly, the titration experiments were repeated for the tri-PBS quenched by the penta-OCP. The shift to Q1 state as the relative amount of OCP is reduced can be clearly observed for both types of phycobilisomes.

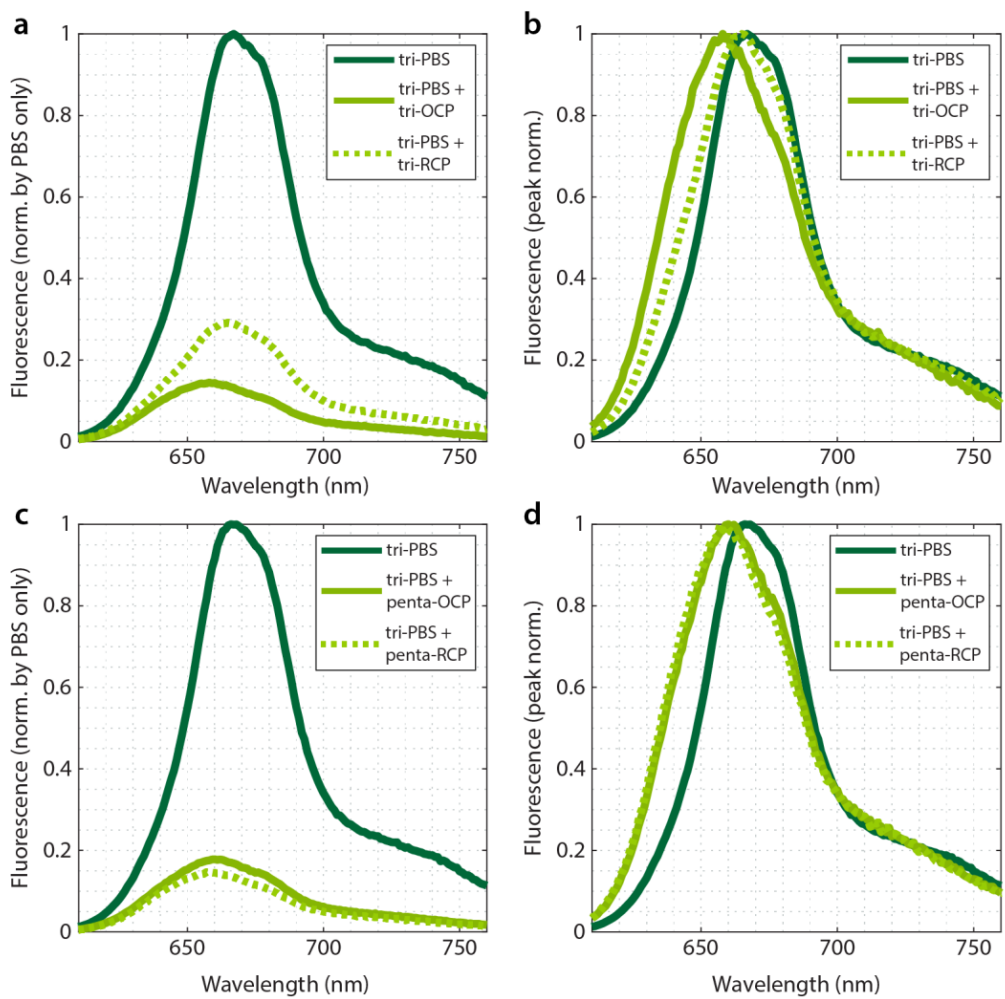

**Fig. S19. Bulk quenching of tri-PBS**

Bulk emission spectra of tri-PBS (dark green) with and without OCP (light green, solid line) or RCP (light green, dashed line) are shown. Activated OCP or RCP (2.5 µM) was added to tri-PBS (5 nM) and incubated for 10 minutes. The sample was excited at 594 nm and the fluorescence emission was measured from 605 to 760 nm. Spectra in (a) and (c) are normalized by the peak of the phycobilisome spectra to visualize the degree of quenching. OCP and RCP from both species can quench the tri-PBS. (a) Upon adding tri-OCP or tri-RCP, the fluorescence brightness decreased to 16% or 31% of the initial unquenched phycobilisome brightness, respectively. (c) Upon adding penta-OCP or penta-RCP, the fluorescence brightness decreased to 20% or 16% of the initial unquenched phycobilisome brightness, respectively. Peak-normalized spectra (b) and (d) show the blue shift of properly quenched phycobilisomes.

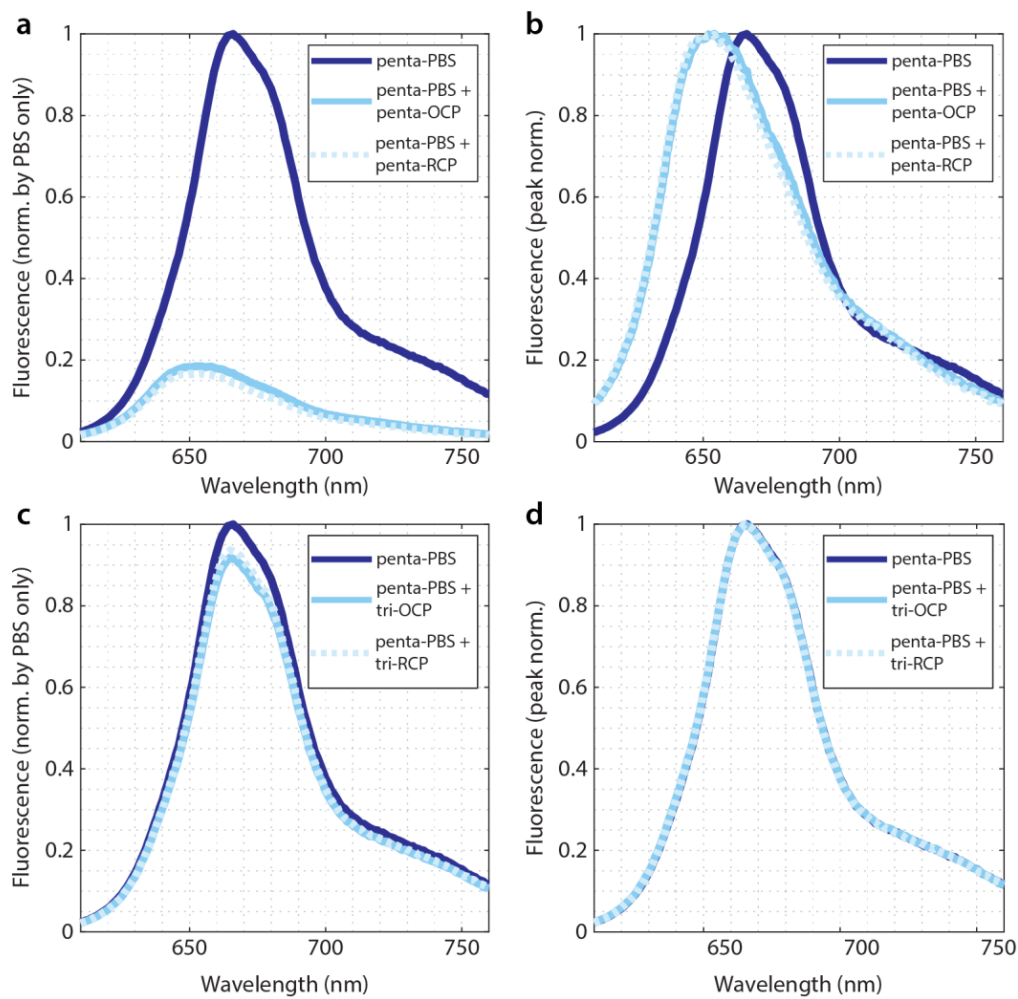

**Fig. S20. Bulk quenching of penta-PBS**

Bulk emission spectra of penta-PBS (dark blue) with and without OCP (light blue, solid line) or RCP (light blue, dashed line). Activated OCP or RCP (2.5 µM) was added to penta-PBS (5 nM) and incubated for 10 minutes. The sample was excited at 594 nm and the fluorescence emission was measured from 605 to 760 nm. Spectra in (a) and (c) are normalized by the peak of the phycobilisome spectra to visualize the degree of quenching. (a) Upon adding penta-OCP or penta-RCP, the fluorescence brightness decreased to 21% or 19% of unquenched PBS fluorescence, respectively. (b) Peak-normalized spectra show the blue shift of properly quenched phycobilisomes. (c, d) The tri-OCP or tri-RCP are unable to quench the penta-PBS.

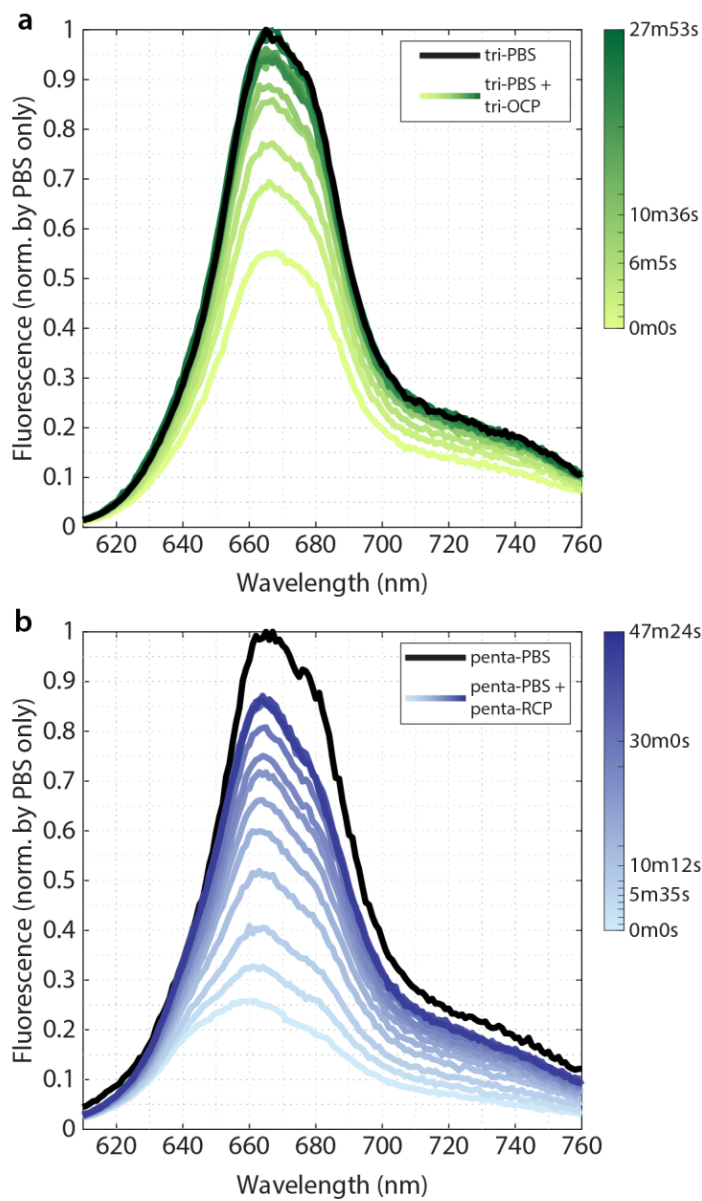

**Fig. S21. RCP unbinds when diluted to low concentrations**

Unlike OCP, RCP unbinds from the phycobilisomes at dilute concentrations as observed by the increase in the fluorescence over time. a) tri-PBS (20 nM) was incubated with tri-RCP (2 µM) for 10 minutes. Then, the mixture was diluted to a final concentration of 0.08 nM PBS and 8 nM RCP. The bulk fluorescence spectrum was repeatedly measured immediately after dilution. To determine the degree of quenching, the fluorescence spectrum of a sample containing the same concentration of PBS without any RCP was also measured (black curve). b) Same procedure was repeated for the penta-PBS quenched with penta-RCP.

| Compartment | No. of compartments | Absorption probability per compartment |
|---|---|---|
| PC640/PC650 | 6 | 0.1097 |
| APC660 | 6 | 0.0507 |
| APC680 | 2 | 0.0187 |

**Fig. S22. Compartmental model for photon-by-photon Monte Carlo simulation of tri-PBS**

Photon-by-photon Monte Carlo simulations were used to simulate quenched phycobilisomes with OCP(s) bound to known binding sites with varying quenching strength. We used the compartmental model given by (3) which places groups of rods and core pigments of the PBS into compartments and uses experimentally determined excitation energy transfer rates to describe the excitation energy flow within the compartments.

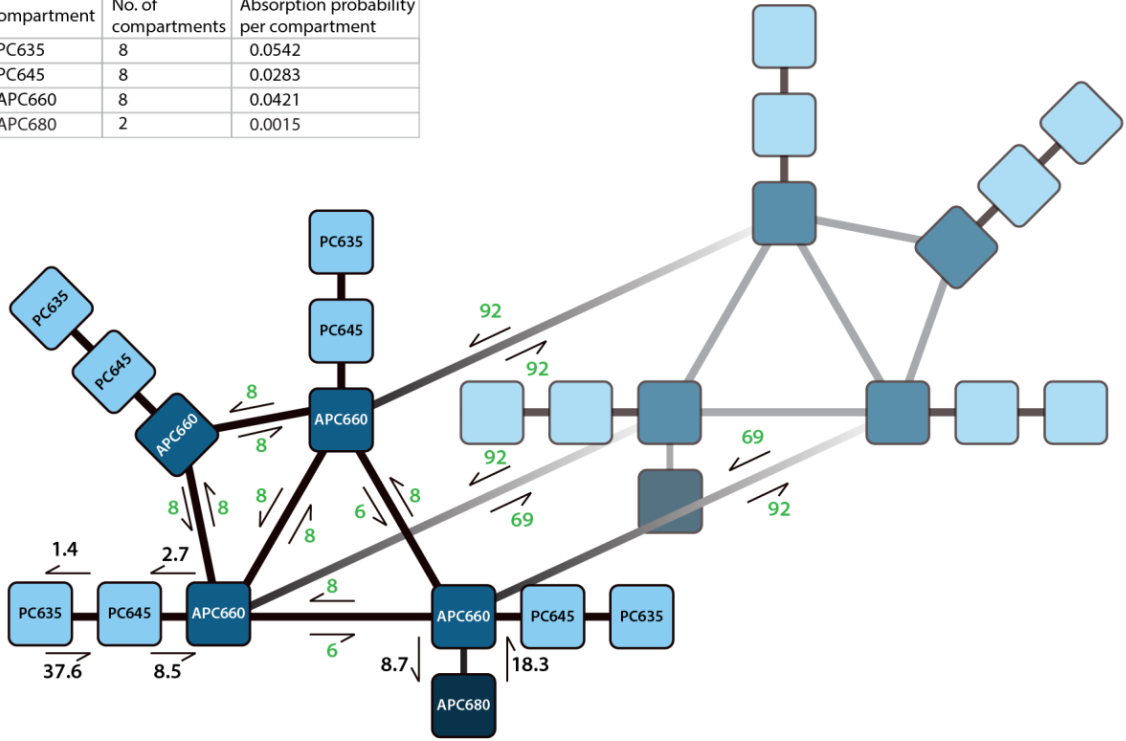

**Fig. S23. Compartmental model for photon-by-photon Monte Carlo simulation of penta-PBS**

Photon-by-photon Monte Carlo simulations were used to simulate quenched phycobilisomes with different number of OCPs bound to different compartments with varying quenching strength. We modified an existing compartmental model for tri-PBS (3, 4) by adding two additional Apc660 compartments as the flanking cylinders in the core with rods attached to them. We used the simplified model for penta-PBS presented by Biswas et al. (5) for the rates between compartments (shown in black). For the compartments within the core, the rates from the tri-PBS model were used (shown in green).

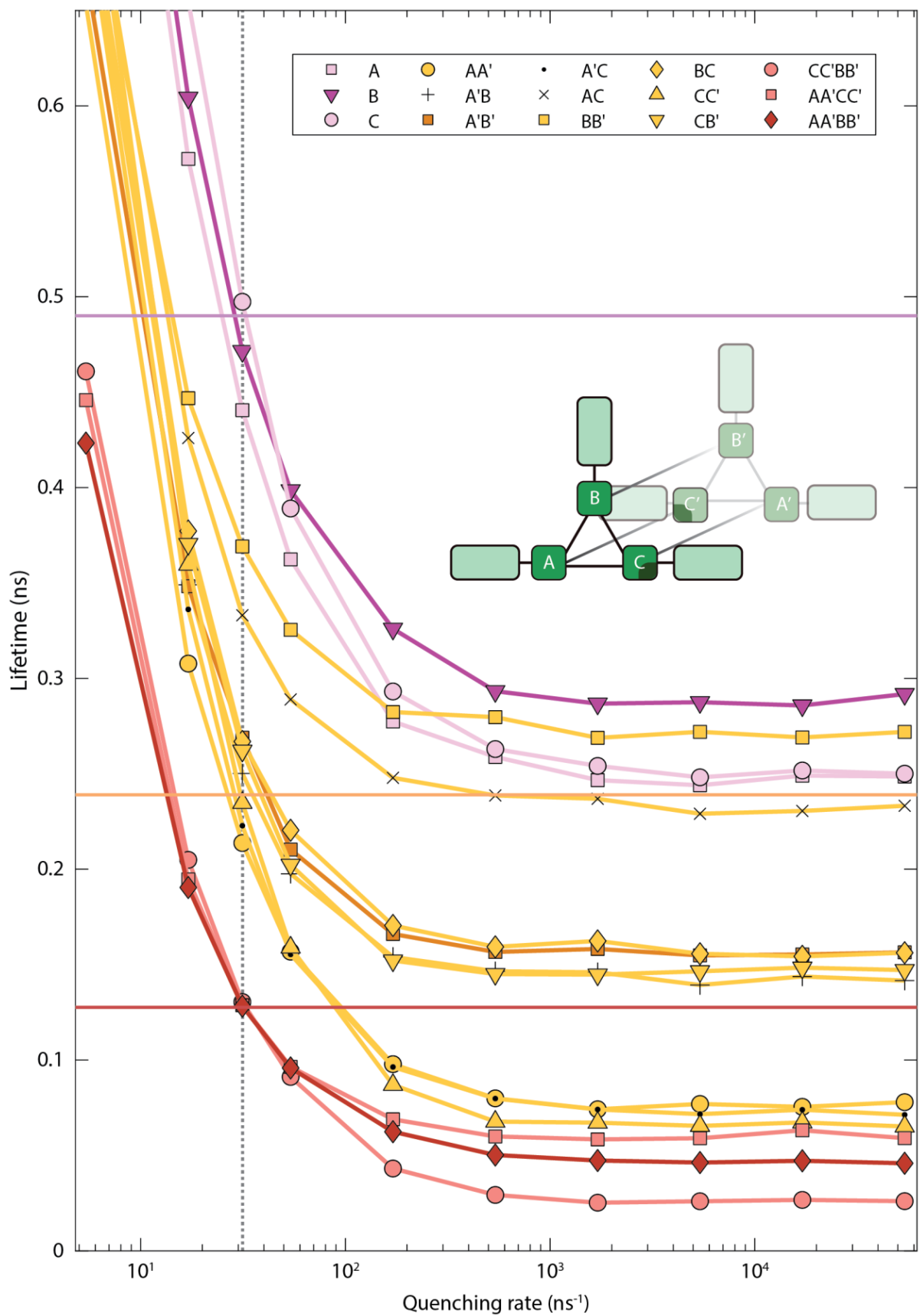

**Fig. S24. (Previous page) Predicted lifetimes for tri-PBS with OCP(s) at various locations.**

Using a compartmental model, one, two, or four OCP(s) with varying quenching strength were placed at different core compartments. At a quenching rate of 31 ns$^{-1}$ (dotted line), two OCP dimers at the known binding sites (AA'BB'; dark red diamond markers) produce the experimentally determined Q2 lifetime (red horizontal line). 2 OCP dimers at other possible sites also produce similar lifetimes (red markers). At the same quenching strength, one OCP dimer at the known binding sites (A'B'; dark green square markers) produces the experimentally determined Q1 lifetime (green horizontal line) and a single OCP monomer at any of the core compartments (purple markers) produces the experimentally determined Qa lifetime (purple horizontal line).

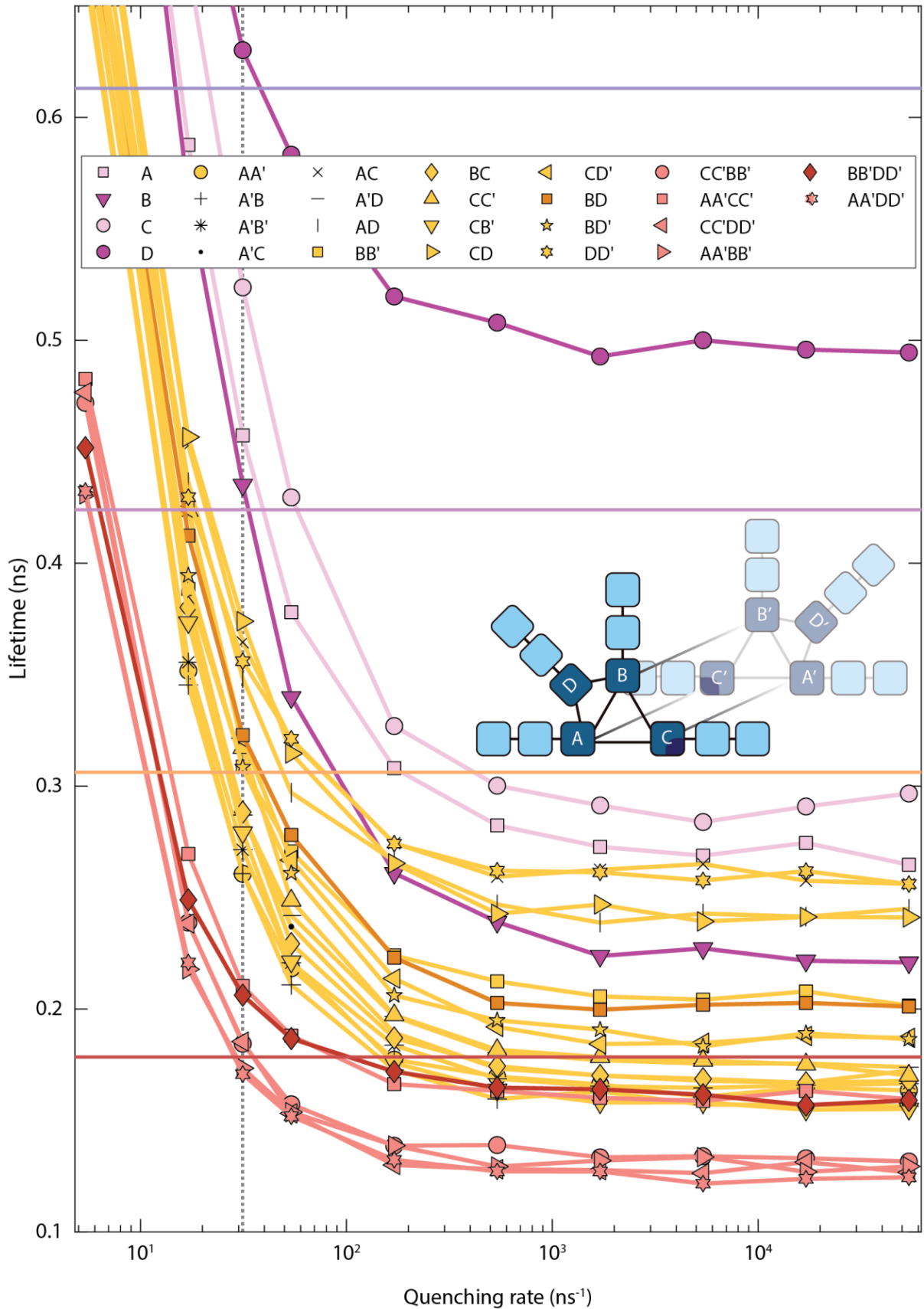

**Fig. S25. (Previous page) Predicted lifetimes for penta-PBS with OCP(s) at various locations.**

Using a compartmental model, one, two, or four OCP(s) with varying quenching strength were placed at different core compartments of the penta-PBS. At a quenching rate of 31 ns$^{-1}$ (dotted line), two OCP dimers are required to achieve the experimentally determined Q2 lifetime (red horizontal line). At the same quenching strength, one OCP dimer could produce the experimentally determined Q1 lifetime (green horizontal line). A single OCP monomer at any of the central core compartments (A, B, or C) could produce the experimentally determined Qa lifetime (purple horizontal line) while an OCP monomer at the flanking core compartment (D) could produce the experimentally determined Qb lifetime (cyan horizontal line).

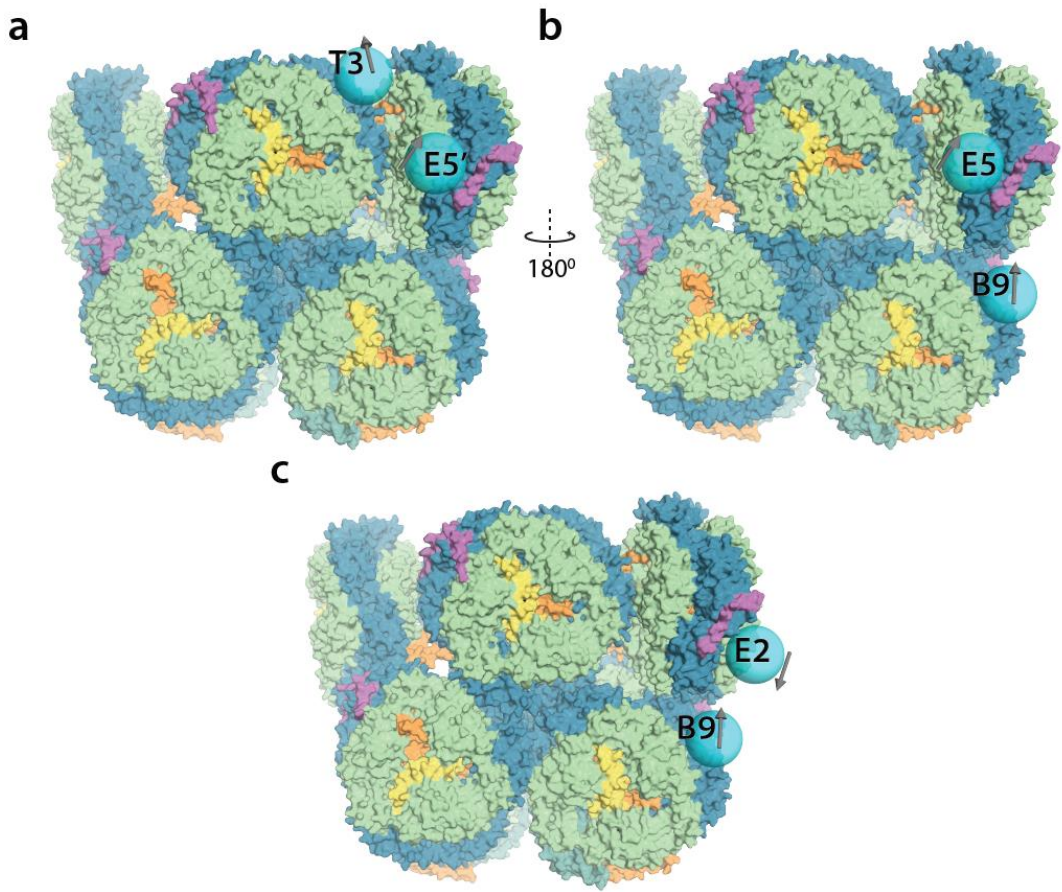

**Fig. S26. Additional potential sites for dimeric OCP^R binding on penta-PBS core**

Pairs of possible binding sites where an OCP^R dimer could feasibly bind are shown as cyan spheres. The arrows on the spheres indicate linker orientation of the aligned NTD of OCP^R. Candidate binding sites were identified by aligning the known binding site on the tri-PBS with the same binding motif (two ApcA and one ApcB) on the penta-PBS. **(a)** Sites T3 and E5' (NTD spacing: 64.8 Å; linker orientation: 68°). **(b)** Sites B9 and E5 (NTD spacing: 73.7 Å; linker orientation: 50°). **(c)** Sites B9 and E2 (NTD spacing: 48.9 Å; linker orientation: 164°).

# Supplementary Tables

**Supplementary Table S1: All available binding sites for the NTD of OCP$^R$ and their availability in the tri-PBS and penta-PBS.**

See **Fig. S4 and S7** for the nomenclature used for the rods and binding sites.

| Site | Availability in tri-PBS | Availability in penta-PBS |
|---|---|---|
| T1 | Available (confirmed binding site) | Available |
| T2 | Blocked by B' | Blocked by B' |
| T3 | Available | Available |
| T4 | Blocked by R3 | Blocked by R3 |
| T5 | Blocked by B | Blocked by B |
| T6 | Partially blocked by R2' | Blocked by E' |
| B1 | Partially blocked by R1 | Available |
| B2 | Bottom site | Bottom site |
| B3 | Blocked by T | Blocked by T |
| B4 | Available (confirmed binding site) | Blocked by E |
| B5 | Bottom site | Bottom site |
| B6 | Blocked by B' | Blocked by B' |
| B7 | Blocked by R1 | Blocked by R1 |
| B8 | Blocked by T | Blocked by T |
| B9 | Available | Available |
| B10 | Blocked by B' | Blocked by B' |
| E1 |  | Blocked by R3 and Re |
| E2 |  | Available |
| E3 |  | Available |
| E4 |  | Blocked by R3 |
| E5 |  | Available |
| E6 |  | Blocked by B |

**Supplementary Table S2: RMSD of all possible OCP binding sites aligned with the known binding site T1 on the top cylinder of the tri-PBS.**

**Note that the T1-T1 RMSD for tri-PBS in line 1 of the table is not zero because the refined structure showing OCP[R] bound to the top core cylinder (PDB: 8TPJ) was used to align with all possible binding sites, including T1. This serves as a baseline to illustrate that all sites will have nonzero RMSD.

Entries for unavailable sites are greyed out. Entries for available sites are colored in green (tri-PBS) or blue (penta-PBS). Entries for repeated sites (symmetric C2) are colorless.

| Binding Site | RMSD in tri-PBS (Å) | RMSD in penta-PBS (Å) |
|---|---|---|
| T1 | 0.418** | 0.678 |
| T2 | 3.273 | 2.975 |
| T3 | 0.685 | 0.891 |
| T4 | 0.690 | 0.757 |
| T5 | 3.160 | 2.475 |
| T6 | 0.671 | 0.954 |
| T1' | 0.407 | 0.712 |
| T2' | 3.198 | 2.839 |
| T3' | 0.791 | 0.881 |
| T4' | 0.590 | 0.752 |
| T5' | 3.213 | 2.586 |
| T6' | 0.808 | 0.865 |
| B1 | 0.628 | 0.954 |
| B2 | 0.549 | 0.744 |
| B3 | 3.711 | 3.794 |
| B4 | 0.800 | 0.918 |
| B5 | 0.664 | 0.897 |
| B6 | 3.332 | 3.764 |
| B7 | 2.101 | 1.823 |
| B8 | 1.333 | 1.361 |
| B9 | 2.206 | 2.246 |
| B10 | 1.369 | 1.389 |
| B1' | 0.677 | 0.999 |
| B2' | 0.602 | 0.703 |
| B3' | 3.588 | 3.792 |
| B4' | 0.740 | 0.943 |
| B5' | 0.717 | 0.836 |
| B6' | 3.208 | 3.806 |
| B7' | 2.090 | 1.972 |
| B8' | 1.392 | 1.353 |
| B9' | 2.220 | 2.267 |
| B10' | 1.412 | 1.360 |
| E1 |  | 1.341 |
| E2 |  | 1.212 |
| E3 |  | 1.206 |
| E4 |  | 1.164 |
| E5 |  | 1.166 |
| E6 |  | 1.174 |
| E1' |  | 1.289 |
| E2' |  | 1.160 |
| E3' |  | 1.190 |
| E4' |  | 1.235 |
| E5' |  | 1.209 |
| E6' |  | 1.164 |

**Supplementary Table S3: Distance and angle between known OCP binding site locations on the tri-PBS.**

Using the known OCP-bound structure of tri-PBS (PDB: 7SC9 (8)), we calculated the distance between the centers of mass of the NTD and the linker orientation of OCP dimers.

| Binding Site 1 | Binding Site 2 | NTD-to-NTD distance (Å) | Cos (Θ) |
|---|---|---|---|
| T1 | B4 | 84.8 | -0.83 |
| T1' | B4' | 84.3 | -0.83 |

**Supplementary Table S4: Distance and angle between pairs of accessible OCP-binding sites on penta-PBS.**

For each pair of accessible binding site on the penta-PBS, the OCP$^R$(NTD) was aligned to the sites and the distance between the center of mass of the NTD was calculated. The cos(θ) value for the pair of linker orientation near the NTD was also calculated to compare the relative orientation of the linkers among different binding sites. Candidate binding pairs are highlighted in blue.

| Binding Site 1 | Binding Site 2 | NTD-to-NTD distance (Å) | Cos(θ) | Reasoning |
|---|---|---|---|---|
| T1 | T3 | 100.9 | -0.21 | Same compartment |
|  | B1 | 96.7 | 0.99 | Linkers too aligned |
|  | B9 | 155.3 | -0.51 | Too far |
|  | E2 | 155.4 | 0.26 | Too far |
|  | E3 | 79.5 | -0.89 |  |
|  | E5 | 145.2 | 0.15 | Too far |
| T3 | B1 | 183.5 | -0.16 | Too far |
|  | B9 | 213.9 | 0.73 | Too far |
|  | E2' | 99.4 | -0.83 | Too far |
|  | E3' | 116.3 | 0.72 | Too far |
|  | E5' | 64.8 | 0.38 |  |
| B1 | B9 | 97.6 | -0.56 | Same compartment |
|  | E2 | 110.8 | 0.31 | Too far |
|  | E3 | 24.2 | -0.94 | Too close |
|  | E5 | 133.1 | 0.17 | Too far |
| B9 | E2 | 48.9 | -0.96 |  |
|  | E3 | 95.9 | 0.48 | Steric clash |
|  | E5 | 73.7 | 0.63 |  |
| E2 | E3 | 100.4 | -0.24 | Same compartment |
|  | E5 | 47.1 | -0.77 | Same compartment |
| E3 | E5 | 117.5 | -0.36 | Same compartment |

# Supplementary References